\documentclass{svjour3}
\usepackage[T1]{fontenc}
\usepackage[latin9]{inputenc}
\usepackage{url}
\usepackage{amsbsy,amsmath,amstext}
\usepackage{graphicx,subfigure}
\usepackage{esint}
\usepackage{subfig}
\newcommand{\be} {\begin{equation}}
\newcommand{\ee} {\end{equation}}
\newcommand{\bea}{\begin{eqnarray}}
\newcommand{\eea}{\end{eqnarray}}
\makeatletter
\RequirePackage{fix-cm}
\smartqed
\journalname{Celestial Mechanics and Dynamical Astronomy}
\makeatother
\begin{document}

\title{The Resonance Overlap and Hill Stability Criteria Revisited}
\author{X.S. Ramos \and J.A. Correa-Otto \and C. Beaug\'e}
\institute{X.S. Ramos \at \emph{Instituto de Astronom\'{\i}a Te\'orica y Experimental (IATE), Observatorio Astron\'omico, Universidad Nacional de C\'ordoba, Argentina} \\
\email{xramos@oac.unc.edu.ar} \\
\and J.A. Correa-Otto \at \emph{Complejo Astron\'omico El Leoncito (CASLEO-CONICET), San Juan, Argentina}\\
\email{jcorrea@casleo.gov.ar}\\
\and C. Beaug\'e \at \emph{Instituto de Astronom\'{\i}a Te\'orica y Experimental (IATE), Observatorio Astron\'omico, Universidad Nacional de C\'ordoba, Argentina} \\
\email{beauge@oac.uncor.edu}}
\titlerunning{The resonance overlap and Hill stability criteria revisited}
\authorrunning{X. S. Ramos et al.}
\date{Received: date /Accepted: date}
\maketitle
\begin{abstract}
We review the orbital stability of the planar circular restricted three-body problem, in the case of massless particles initially located between both massive bodies. We present new estimates of the resonance overlap criterion and the Hill stability limit, and compare their predictions with detailed dynamical maps constructed with N-body simulations. We show that the boundary between (Hill) stable and unstable orbits is not smooth but characterized by a rich structure generated by the superposition of different mean-motion resonances which does not allow for a simple global expression for stability. 

We propose that, for a given perturbing mass $m_1$ and initial eccentricity $e$, there are actually two critical values of the semimajor axis. All values $a < a_{\rm Hill}$ are Hill-stable, while all values $a > a_{\rm unstable}$ are unstable in the Hill sense. The first limit is given by the Hill-stability criterion and is a function of the eccentricity. The second limit is virtually insensitive to the initial eccentricity, and closely resembles a new resonance overlap condition (for circular orbits) developed in terms of the intersection between first and second-order mean-motion resonances.
\end{abstract}


\section{Introduction\label{intro}}

The question of orbital stability in the circular restricted three-body problem (CR3BP) is a long-standing and complex problem. Our particular interest can be summarized in the following manner. Assume a massless particle (e.g. asteroid) orbiting a central star with mass $m_0$ and perturbed by a massive planet $m_1$. We will denote by $a$ the osculating semimajor axis of the particle, $e$ its eccentricity, $\lambda $ the mean longitude and $\varpi $ the longitude of perihelion. Elements with subindex $1$ correspond to the perturber, whose orbit is considered circular (i.e. $e_1=0$) and exterior to that of the particle ($a < a_1$). We will also assume that all motion is restricted to the plane. Under these considerations, given a certain eccentricity $e$ for the particle, and fixing the angles at a certain value, what is the critical semimajor axis that separates the domains of stable and unstable motion? 

This problem has been addressed by different methods, depending on the type of stability under consideration. The simplest is the so-called Hill Stability, in which an initial condition is said to be stable if its Jacobi constant $C_J$ is larger than the value $C_{L_1}$ it acquires at the $L_1$ Euler-Lagrange point of the system. The particle will then be trapped within a Hill zero-velocity region that excludes the position of $m_1$. The trajectory will never be able to cross the orbit of the perturber and will therefore remain bounded. 

The outcome of initial conditions that do not satisfy the Hill Stability criterion is not obvious. 
While the condition $C_J > C_{L_1}$ is sufficient for stability, it is not necessary. It is possible to find solutions that do not comply with this inequality, but are nevertheless stable, at least for times of the order of Gyrs (e.g. Gladman 1993). As we will show in this paper, some of these initial conditions lie within the librational domain of mean-motion resonances, but others are non-resonant.

A second estimator, this time of orbital instability, is the Resonance Overlap criterion, based on the work of Chirikov (1979) and first applied to the three-body problem by Wisdom (1980). As its name indicates, it postulates that global chaos (and therefore orbital instability) is triggered by the overlap of adjacent mean-motion resonances (MMRs). Wisdom concentrated on the case of circular orbits ($e=0$) and first-order commensurabilities. His overlap criterion stipulates that instability will occur whenever the distance between two consecutive MMRs is smaller or equal to the sum of their libration widths. 

Using an analytical model which would later be known as the Second Fundamental Model for Resonance (SFMR, Henrard and Lema\^{\i}tre 1983) and assuming that the sum of the libration widths could be approximated by twice the size of the inner separatrix of the commensurability farther from the perturber, Wisdom (1980) estimated that the {\it averaged} semimajor axis $a^*_{\rm overlap}$ leading to overlap may be approximated by:
\be
a^*_{\rm overlap} = a_1 \biggl[ 1 - D \biggl( \frac{m_1}{m_0} \biggr)^{2/7} \biggr]
\label{eq1}
\ee
where $D = 1.3$ is a constant. This expression assumes initial conditions such that the resonant angles $\sigma$ are zero, where the libration region of first-order resonances is at its maximum. 

Malhotra (1998) and Deck et al. (2013) presented new calculations of the overlap limit, using a similar analytical model description for the resonant Hamiltonian but with different degrees of approximations. Deck et al. (2013), for example, took into account that adjacent resonances do not have the same libration half-widths, while both papers considered slightly different expressions for the resonance half-width. Their results show the same functional form in terms of the planetary mass, although with different numerical coefficients: $D=1.4$ in the case of Malhotra (1998) and $D=1.46$ for Deck et al. (2013). 

Duncan et al. (1989) avoided analytical methods and attacked the problem using numerical simulations with a symplectic mapping. Their results once again yielded the same dependence on $m_1/m_0$, but with a larger coefficient: $D=1.49$. Consequently, the so-called $(m_1/m_0)^{2/7}$-law appears extremely robust to the modeling of the problem, although the numerical factor shows a significant spread and is less reliable.

The case of eccentric orbits ($e > 0$) is more problematic. While numerical experiments by Quillen and Faber (2006) seem to indicate that there should not be any significant different in the resonant overlap limit for moderate eccentricities, analytical studies of Mustill and Wyatt (2012) and Deck et al. (2013) point in the opposite direction. Both papers predict that even for low values of the eccentricity the overlap distance occurs much farther from the planets and, more surprisingly, the dependence with the planetary mass changes to $(m_1/m_0)^{1/5}$. 

In this paper we revisit the resonance overlap criterion and its relation with the Hill stability limit. For the overlap calculations, we once again make use of the second fundamental model of resonance but present a new approach to the calculation of the overlap condition. The main difference with respect to previous studies is twofold. First, we show that for circular orbits there is no outer separatrix, and thus the basic idea of equating the resonance separation to the sum of the separatrix widths is not obvious. 

Second, we show that second-order mean-motion resonances are also important in determining the instability limit, even for initial conditions where their libration widths is minimum. Curiously, this new overlap criterion is very similar as obtained with the classical model, although systematically closer to the planet. We also show that our new overlap criterion can be used as a empirical estimate for the critical semimajor axis (which we denote by $a_{\rm unstable}$) that marks the beginning of the chaotic sea and completely unstable orbits. 

For the Hill Stability, we present simple expressions to calculate this limit for any initial condition, and compare their predictions with the resonance overlap. We show that both criteria are different but complementary; while $a_{\rm unstable}$ marks the lower limit for global orbital instability, Hill stability marks the end of the stable and bounded motion. In between lies a rugged region of the phase space dominated by complex resonant structures where both stable and unstable orbits may be found.

Finally, we analyze the case of eccentric orbits. We find that even for $e \sim 0.4$ the expression for $a_{\rm unstable}$ deduced for circular orbits is a very good indication of the global chaotic domain. The Hill Stability limit, however, is very sensitive to this parameter, and the transition region between both regimes grows with the eccentricity. Therefore a single criterion cannot be proposed as a unique law separating stable from unstable motion, especially for eccentric orbits.

\section{The Resonant Hamiltonian\label{model}}

\subsection{Delaunay and Resonant Canonical Variables}

Since we will be working within the Hamiltonian formalism, we will introduce the usual modified Delaunay canonical variables:
\bea
L = \sqrt{\mu a} \hspace*{2.57cm} &;& \hspace*{0.5cm} \lambda \nonumber \\
S = \sqrt{\mu a}(1-\sqrt{1-e^2}) \hspace*{0.5cm} &;&\hspace*{0.2cm} -\varpi  \\
\Lambda    \hspace*{3.7cm} &;&\hspace*{0.5cm} \lambda_1 \nonumber
\label{eq2}
\eea
where $\mu ={\cal G} m_0$ and ${\cal G}$ denotes the gravitational constant. Since the longitude of pericenter of the planet is constant, it does not appear as a variable of the dynamical system. $\Lambda$ is the canonical momentum associated to $\lambda_1$, and its value is not known a priori.
The Hamiltonian of the system in the extended phase space can be written as:
\be
F(L,S,\Lambda,\lambda,\varpi,\lambda_1) = -\frac{\mu^2}{2L^2} + n_1 \Lambda - R(L,S,\lambda,\varpi,\lambda_1;a_1,e_1,\varpi_1) ,
\label{eq3}
\ee
where $n_1$ is the mean motion of the perturber and $R$ represents the disturbing function due to the gravitational perturbations of $m_1$. 

Let us now suppose that the massless particle lies in the vicinity of a generic $(p+q)/p$ mean-motion resonance with the perturber, such that 
\be
(p+q) n_1 - p n \simeq 0,
\label{eq4}
\ee
where $n$ is the mean motion of the particle and both $p$ and $q$ are positive integers. It is then convenient to introduce a set of resonant canonical variables $(S,N,\Lambda',\sigma,\nu,Q)$ which are related to the Delaunay variables through
\be
\begin{split}
S \hspace*{3.05cm} &; \hspace*{0.5cm} q\sigma = \;\; (p+q)\lambda _1-p\lambda -q\varpi  \\
N = S - L - \Lambda \hspace*{1.03cm} &; \hspace*{0.5cm} q\nu
=-(p+q)\lambda _1+p\lambda +q\varpi _1  \\
\Lambda' = p \Lambda + (p+q) L \hspace*{0.5cm} &; \hspace*{0.5cm} qQ = \lambda - \lambda_1  ,
\end{split}
\label{eq5}
\ee
where $qQ$ is the synodic angle. The inverse transformation can be obtained easily 
after some cumbersome algebraic manipulations, and yields:
\be
\begin{split}
S \hspace*{4.5cm} &; \hspace*{0.5cm} M = \sigma + (p+q) Q  \\
L = \frac{p}{q}(N-S) + \frac{1}{q}\Lambda' \hspace*{1.6cm} &; \hspace*{0.5cm} M_1 = -\nu + pQ \\
\Lambda = -\frac{(p+q)}{q} (N-S) - \frac{1}{q}\Lambda' \hspace*{0.5cm} &; \hspace*{0.5cm} \varpi = \varpi_1 - \sigma - \nu ,
\end{split}
\label{eq6}
\ee
where $M$ and $M_1$ are the mean anomalies. Since the transformation $(L,S,\Lambda,\lambda,\varpi,\lambda_1) \rightarrow (S,N,\Lambda',\sigma,\nu,Q)$ is canonical, the Hamiltonian of the extended phase space is preserved.

\subsection{Averaging over the Synodic Angle}

In the vicinity of a mean-motion resonance, both $\sigma$ and $\nu$ are slowly varying angles (i.e. long-period variables) while $Q$ has a high frequency of the order of the orbital periods of the bodies. Moreover, the amplitude of the short-period variations are usually much smaller than their resonant and secular counterparts, and therefore have little effect on the long-term evolution of the system. It is thus useful to average the Hamiltonian with respect to $Q$ and eliminate the short-period variations. 

The averaging is usually accomplished through a perturbation technique such as Hori's method (Hori 1966, see also Ferraz-Mello 2007). Basically, we search for a Lie-type canonical transformation 
\be
B(S^*,N^*,\Lambda'^*,\sigma^*,\nu^*,Q^*) : (S,N,\Lambda',\sigma,\nu,Q) \rightarrow (S^*,N^*,\Lambda'^*,\sigma^*,\nu^*,Q^*)
\label{eq7}
\ee
to new (primed) variables such that the new Hamiltonian $F^*$ is independent of $Q^*$. Although the construction of $B$ is complicated when extended to high orders of the small parameter (here the ratio $m_1/m_0$), when restricted to first order it can be simply thought as the definite integral of $F$ over $Q$ in the interval $[0,2\pi]$. The same procedure can also be performed numerically, yielding a semi-analytical expression for the averaged Hamiltonian (e.g. Moons and Morbidelli 1993, Beaug\'e 1994). 

Whichever the method adopted, we obtain a new function $F^*(S^*,N^*,\Delta'^*,\sigma^*,\nu^*)$ which is cyclic in $Q^*$. Consequently, the corresponding canonical momenta $\Lambda'^*$ is an integral of motion of the system. Notice from the transformations (\ref{eq6}) that $\Lambda'^*$ just appears as an additive constant in the relation between the momenta. So, independently of the initial conditions, we can just choose $\Lambda'^* = 0$ and simplify both the Hamiltonian and the canonical transformations. 

In the averaged variables, the momenta in the Delaunay and resonant sets are related via:
\be
L^* = \frac{p}{q} (N^* - S^*)  \hspace*{0.5cm} ; \hspace*{0.5cm} 
\Lambda^* = -\frac{(p+q)}{q} (N^* - S^*) ,
\label{eq8}
\ee
and the averaged Hamiltonian $F^*$ can be written as:
\be
F^*(S^*,N^*,\sigma^*,\nu^*) = -\frac{\mu^2q^2}{2p^2} (N^*-S^*)^{-2} - \frac{(p+q)}{q}n_1 (N^*-S^*)  - R^*(S^*,N^*,\sigma^*,\nu^*),
\label{eq9}
\ee
where $R^*$ is now the disturbing function averaged over short-period terms. We define the value of $L^*$ at exact resonance as:
\be
\frac{\mu^2}{{{L^*}_{\rm res}}^3} = n^*_{\rm res} = \frac{(p+q)}{p} n_1 ,
\label{eq10}
\ee
and consider only initial conditions close to exact resonance. We can then expand the unperturbed part of $F^*$ as a Taylor series around ${{L^*}_{\rm res}}$. Retaining only second-order terms, we can write:
\be
F^*(S^*,N^*,\sigma^*,\nu^*) \simeq - A_0 {(N^* - S^*)}^2 + A_1 (N^* - S^*) - R^*(S^*,N^*,\sigma^*,\nu^*),
\label{eq11}
\ee
where $A_i$ are positive constants that depends only on $p,q$ and the masses:
\be
A_0 = \frac{3p^2\mu^2}{2q^2{{L^*}_{\rm res}}^4} \hspace*{0.5cm} ; \hspace*{0.5cm}
                       A_1 = \frac{3p\mu^2}{q{{L^*}_{\rm res}}^3} . 
\label{eq12}
\ee

\subsection{The Circular Problem}

We now consider the case where $m_1$ moves in a circular orbit (i.e. $e_1=0$). The disturbing function is only function of $\sigma$, the auxiliary angle $\nu$ is cyclic and the associated momentum $N^*$ is an integral of motion. From equations (\ref{eq5}) and (\ref{eq8}) we can write:
\be
N^* = S^* + \frac{q}{p}L^* = \sqrt{\mu a^*} \biggl( \frac{(p+q)}{p} - \sqrt{1-{e^*}^2} \biggr) 
    = const.
\label{eq13}
\ee
This implies that, given any initial values of the mean semimajor axis and eccentricity, their time evolution will preserve the value of $N^*$. Both orbital elements are thus not independent, but coupled. The complete Hamiltonian can now be written as:
\be
F^*(S^*,\sigma^*;N^*) = -\frac{\mu^2q^2}{2p^2} (N^*-S^*)^{-2} - \frac{(p+q)}{q}n_1 (N^*-S^*) 
   - R^*(S^*,\sigma^*;N^*) ,
\label{eq14}
\ee
which is a single degree-of-freedom system parametrized by $N^*$. 

We now turn our attention to the expression of $R^*(S^*,\sigma;N^*)$ adopted for most analytical resonance models. From the Laplace expansion of the disturbing function, we will retain only the lowest-order secular and resonant terms, and thus write:
\be
R^* = \frac{{\cal G} m_1}{a_1} \biggl( {\hat g}_{0,0}(\alpha^*) + 
               {\hat g}_{0,1}(\alpha^*) {e^*}^2 +
               {\hat g}_{1,0}(\alpha^*) e^* \cos{\sigma^*} \biggr) . 
\label{eq15}
\ee
In the case of first-order resonances, the expressions for the coefficients can be found in Brouwer and Clemence (1961) or Murray and Dermott (1999), and read:
\be
\begin{split}
{\hat g}_{0,0}(\alpha^*) &=  \;\;\; \frac{1}{2} b^{(0)}_{1/2} (\alpha^*) \\
{\hat g}_{0,1}(\alpha^*) &=  \;\;\; \frac{1}{8} \biggl[ 2 \alpha^* D_\alpha + {\alpha^*}^2 D_\alpha^2 \biggr] b^{(0)}_{1/2} (\alpha^*) \\
{\hat g}_{1,0}(\alpha^*) &= -\frac{1}{2} \biggl[ 2(p+q) + \alpha^* D_\alpha\biggr] b^{(p+q)}_{1/2}(\alpha^*) ,
\end{split}
\label{eq16}
\ee
where $D_\alpha \equiv d/d\alpha$ is the differential operator, and $b^{(j)}_{1/2}$ are Laplace coefficients. 

We will make two additional approximations. First, we will evaluate all coefficients at the exact resonance $\alpha^*_{\rm res} = a^*_{\rm res}/a_1$. Since the perturbation is small compared to the unperturbed Hamiltonian, and we are only interested in the vicinity of the exact resonance, then the error committed here is not significant. Second, we will approximate the eccentricities by:
\be
e^* \simeq \sqrt{ \frac{2S}{L^*} } \simeq \sqrt{ \frac{2S}{L^*_{\rm res}} } .
\label{eq17}
\ee
The same arguments mentioned before are valid here, and again the error generated by this approximation is not relevant, at least up to eccentricities of the order of $e \sim 0.5$. 

Introducing these simplifications into (\ref{eq14}), we can write the complete averaged resonant Hamiltonian for the circular problem as:
\be
F^*(S^*,\sigma^*;N^*) = -A_0 {(N^* - S^*)}^2 + A_1 (N^* - S^*) - C_1 S^* - C_2 \sqrt{2S^*} \cos{\sigma^*} , 
\label{eq18}
\ee
where we have dropped constant terms, and the new coefficients are given by 
\be
C_1 = \frac{{\cal G}m_1}{a_1 {L^*}_{\rm res}} {\hat g}_{0,1}(\alpha_{\rm res}^*) \hspace*{0.5cm} ; \hspace*{0.5cm}
C_2 = \frac{{\cal G}m_1}{a_1 \sqrt{{L^*}_{\rm res}}} {\hat g}_{1,0}(\alpha_{\rm res}^*) .
\label{eq19}
\ee
Expression (\ref{eq18}) constitutes a very simple analytical model for mean-motion resonances in the circular restricted three-body problem and, apart from the Taylor expansion of the unperturbed part, is identical to the Second Fundamental Model of Resonance (Henrard and Lema\^{\i}tre 1983).

\subsection{Fixed Points and Separatrix}

For first-order ($q=1$) resonances, all fixed points are located in either $\sigma^*=0$ or $\sigma^*=\pi$, and parametrized by the value of $N^*$. If this parameter is less than a critical value 
\be
N^*_c = \frac{1}{2A_0} \biggl[ A_1 + C_1 + \biggl( \frac{27}{4} A_0 C_2^2\biggr)^{1/3} \biggr] ,
\label{eq20}
\ee
the system contains a single (stable) fixed point at $\sigma^*=0$. Conversely, if $N^* > N_c$, the Hamiltonian $F^*$ contains three fixed points: two centers (one located at $\sigma^*=0$ and a second one at $\sigma^*=\pi$) plus one unstable point at $\sigma^*=\pi$. The corresponding values of the momentum $S^*$ can be calculated analytically solving the equations of motion. The results, given in complex trigonometric and hyperbolic functions, can then be converted back to $a^*$ and $e^*$. 

\begin{figure}[t!]
\centering
\includegraphics*[width=12cm]{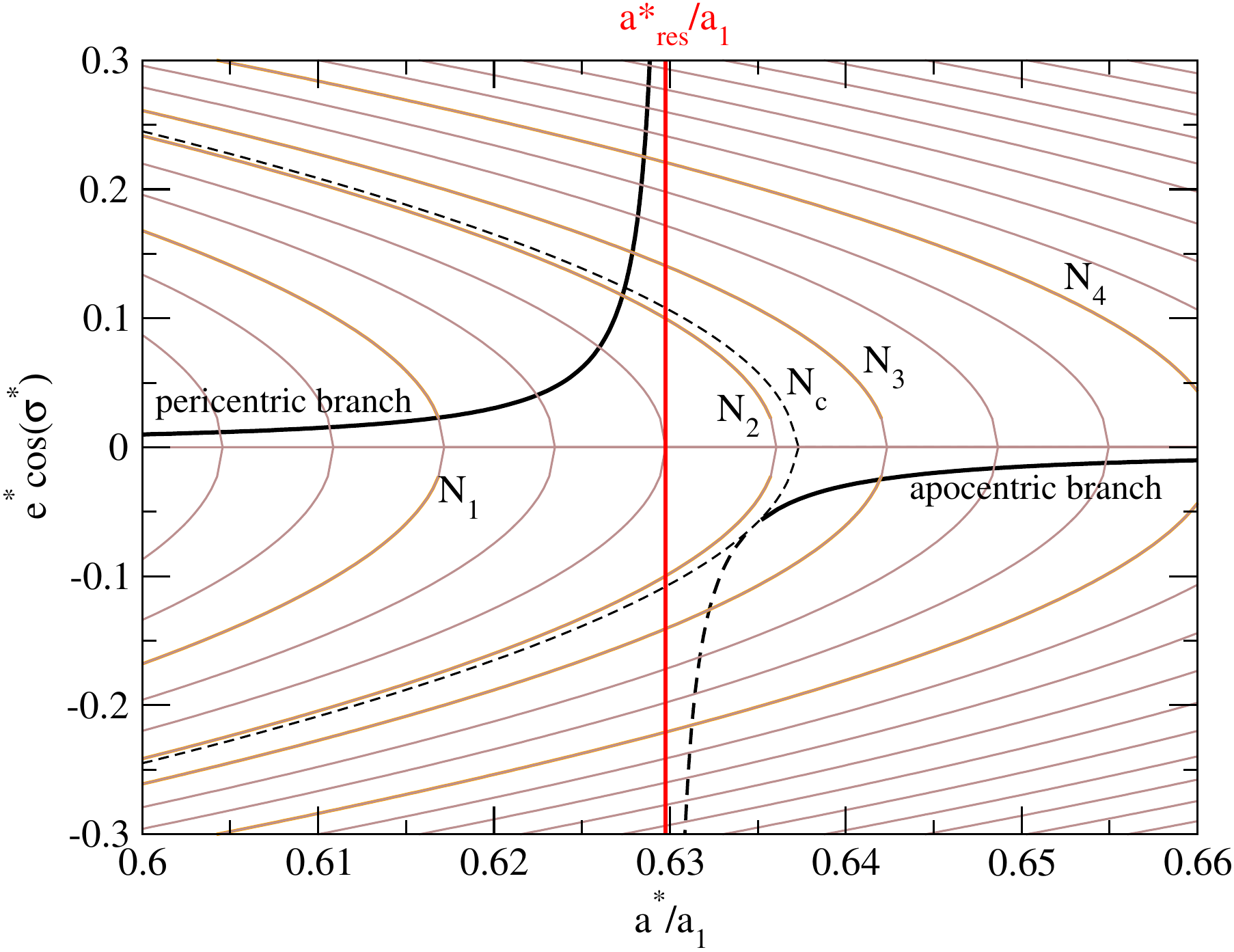}  
\caption{Broad black curves show the families of fixed points for the 2/1 MMR in the $(a^*/a_1,e^*)$ plane, considering Jupiter as the perturber. The top half-plane (positive values of $e^*\cos{\sigma^*}$) correspond to $\sigma^*=0$, while the bottom half-plane corresponds to $\sigma^*=\pi$. The position of the nominal resonant mean semimajor axis is marked by a broad red vertical line. The gray curves correspond to different values of $N^*=const$ (with $N_1 < \ldots < N_4$). the critical value $N^*_c$ is marked by a light black dashed curve.}
\label{fig1}
\end{figure}

Figure \ref{fig1} plots the families of fixed points for the 2/1 MMR, adopting Jupiter (present mass) as the perturber. The gray curves correspond to different values of $N^*=const$, and the critical value $N^*_c$ is marked by a light black dashed curve. The top half-plane (positive values of $e^*\cos{\sigma^*}$) correspond to $\sigma^*=0$, and the fixed points define what is usually known as the {\it pericentric branch}. All fixed points of the pericentric branch are linearly stable. They have low eccentricity far from exact resonance (shown here as a broad vertical red line), but the value of $e^*$ increases as $a^* \rightarrow a^*_{\rm res}$. In no case, however, does the fixed point intersect the nominal location of the resonance, but is always located at smaller values of the semimajor axis. 

The bottom half-plane corresponds to $\sigma^*=\pi$. For low eccentricities, the solutions are again stable, and form what is known as the {\it apocentric branch}. For higher eccentricities (broad dashed curve) the solutions are unstable and correspond to the hyperbolic fixed points from which stem the separatrix of the libration regions. 

Ferraz-Mello (1988) found a simple expression relating the semimajor axis and eccentricity for all fixed points. Although his calculations employed the asymmetric expansion of the disturbing function (Ferraz-Mello 1987), the same procedure can be followed in the case of the SFMR. We begin writing the condition $d \sigma^*/dt = 0$ for the fixed points as
\be
2 A_0 (N^* - S^*) - A_1 - C_1 - \frac{C_2}{\sqrt{2S^*}} \cos{\sigma^*} = 0 ,
\label{eq21}
\ee
where we will consider the resonant angle equal to either zero or $\pi$. Instead of expressing this as an algebraic equation in $S^*$, we recall that $L^* = p(N^*-S^*)/q$ from which we can simply obtain:
\be
\frac{C_2}{\sqrt{2S^*}} \cos{\sigma^*} = 2 A_0 \frac{q}{p}L^* - A_1 - C_1 
                                       = 2 A_0 \frac{q}{p} \biggl( L^* - L^*_c \biggr) , 
\label{eq22}
\ee
where
\be
L^*_c = \frac{p}{q} \frac{A_1 + C_1}{2A_0}
\label{eq23}
\ee
constitutes the equilibrium value of the Delaunay variable. 

We next approximate $\sqrt{2S^*} \simeq e^* \sqrt{L^*} \simeq e^* \sqrt{L^*_{\rm res}}$ and, after some simple substitutions, obtain the eccentricity $e^*$ and the value of $L^*$ for all fixed points:
\be
\frac{1}{e^*} \cos{\sigma^*} = \frac{2A_0 q}{C_2 p} \sqrt{L^*_{\rm res}} \biggl( L^* - L^*_c \biggr). 
\label{eq24}
\ee
Values of $L^* < L^*_c$ give rise to the pericentric branch, while other values generate the apocentric and hyperbolic families. Note, however, that the value of $N^*$ is implicit in this equation, which may complicate the calculation of the libration width. Also, there is no information of the stability of each solution, which must be estimated by additional calculations. Finally, (\ref{eq24}) predicts that the stable and unstable branches are completely symmetrical (or anti-symmetrical) with respect to $L^*_c$. This is not exactly true, but sufficiently accurate for most purposes.

The main advantage of (\ref{eq24}), however, is its simplicity and ease of use. It also shows clearly the hyperbolic natures of the pericentric and apocentric branches and how the locus of fixed points tend to parabolic orbits as we approach exact resonance. 

For $N^* > N^*_c$ we can calculate the borders of the libration region. These will be given by the values of $K = \sqrt{2S^*} \cos{\sigma^*}$, with $\sigma^*=0,\pi$ such that the Hamiltonian coincides with the value at the hyperbolic fixed point. Together with the value of $N^*$ we can then transform them into orbital elements and calculate the values: $(a^*_{\rm in},e^*_{\rm in})$ and $(a^*_{\rm out},e^*_{\rm out})$. The first pair will define the branch of the separatrix separating the libration zone from the inner circulation region or, more correctly, its intersection with the $\sigma^*=0,\pi$ axis, while the second will mark the appearance of the outer circulation domain.

\begin{figure}[t!]
\centering
\includegraphics*[width=12cm]{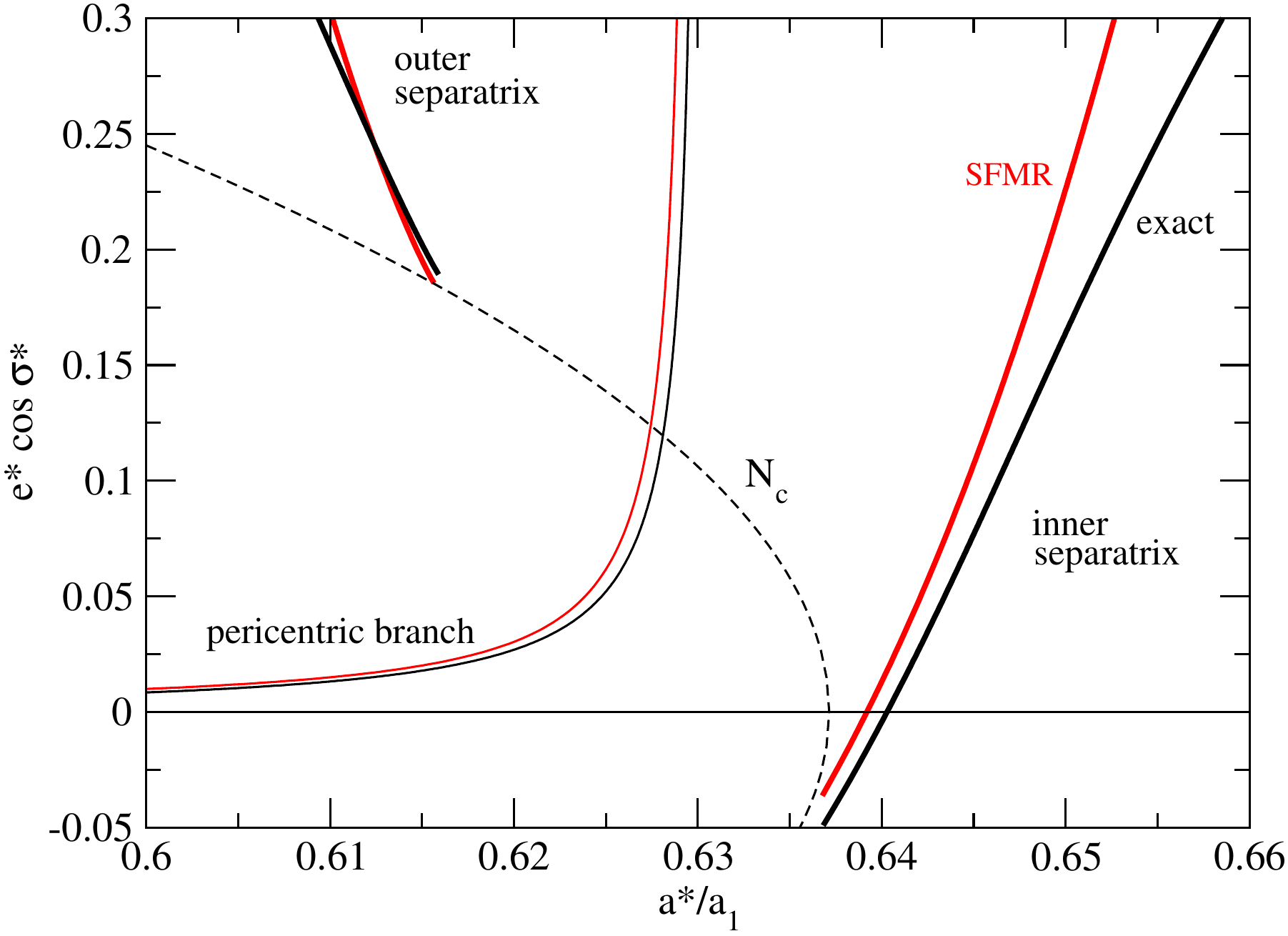}  
\caption{Structure of the 2/1 MMR for the restricted three-body problem with Jupiter in a circular orbit. Thin continuous lines show the locus of stable solutions (pericentric branch), while broad curves mark both branches of the separatrix. The libration region appears only for $N^* \ge N^*_c$ (shown with a dashed black curve). Note that the inner branch of the separatrix extends to values of $e^* \cos{\sigma^*}<0$ which implies $\sigma^*=\pi$.}
\label{fig2}
\end{figure}

Figure \ref{fig2} shows the structure of the MMR with Jupiter as the perturber (present mass). The red curves were calculated using the SFMR Hamiltonian (\ref{eq18}), while the black curves show results using a semi-analytical model for the resonant Hamiltonian in which the averaged disturbing function is evaluated numerically at every point. The SFMR shows very good agreement with the exact calculation, especially for low eccentricities, which justifies the use of the analytical model 
for near-circular orbits. However, the SFMR systematically underestimates the libration width, a datum that will be taken into consideration in later stages of this work. 

To simplify our notation, we will denote by {\it inner separatrix} the locus of points separating the inner circulation region from the librational domain, calculated for all $N^* \ge N_c$. Similarly, we will refer to the boundary between the libration and outer circulation domains as the {\it outer separatrix} of the resonance. Note that the inner branch of the separatrix extends below $e^* \cos{\sigma^*} = 0$, indicating that it is also present for $\sigma^*=\pi$; thus, the limit of the librational domain for circular orbits ($e^*=0$) is not given by the curve $N^*=N^*_c$ but for larger values of the integral of motion.

An important feature of Figure \ref{fig2} that has received little attention is the fact that
the outer separatrix does not extend down to circular orbits. Although for every value of $N^* > N_c$ the system contains both an inner and outer branch, these appear in the $(a^*,e^*)$ plane with
different eccentricities. This is a natural outcome of the fact that both sets $(a^*_{\rm in},e^*_{\rm in})$ and $(a^*_{\rm out},e^*_{\rm out})$ must lead to the same value of $N^*$. 

If we  analyze this figure for a given (fixed) value of the eccentricity, then for $e^* \ge 0.18$ both branches of the separatrix will be present and the librational domain will be contained within. For $e^* < 0.18$, however, there is no outer branch of the separatrix. In other words, for this range of eccentricities all values of the semimajor axis $a^* < a_{\rm res}$ will yield $N^* < N_c$ and will thus correspond to a circulation. There is in fact a libration region, but is much smaller, and limited by the inner separatrix and the value of the semimajor axis such that $N^*=N_c$.

\section{The Classical Resonance Overlap Criterion}

Wisdom (1980) established what may be called the condition for the classical resonance overlap in the CR3BP. Following the ideas of Walker and Ford (1969) and Chirikov (1979), overlap is said to occur when, for a given value of the mean eccentricity $e^*$, the distance $\Delta a^*_{\rm res}$ between adjacent mean-motions resonances is smaller or equal to the sum of their libration widths $\Delta a^*_{\rm sep}$. The same idea was also adopted by later analytical calculations, such as Malhotra (1998) and Deck et al. (2013), although this last
paper improved the model taking into account that the half-widths between neighboring MMRs are
not equal.

All these estimates also employed the SFMR as the resonance model, although with different degrees of approximation of the Hamiltonian, or in the deduction of the resonance width. In many cases these approximations were made in order to obtain results in simple explicit expressions, even though the errors introduced were not adequately checked. 

Finally, resonance overlap was calculated in the representative plane defined by $\sigma^*=\varpi=0$ and evaluated at $e^*=0$. These choices of initial conditions also allowed to consider only the role of first-order commensurabilities. Second-order MMRs have a minimum libration size for $\sigma^*=0$, while the separatrix width of third-order resonances tends to zero for circular orbits. So, focusing on first-order resonances appears a good approximation, especially when studying overlap for circular orbits. 

In this section we once again deduce the resonance overlap limit for circular orbits (i.e. $e^*=0$), following the same assumptions as described above. We will, however, reduce the approximations to a minimum, even if this implies results which are not in explicit analytical expressions.

\subsection{Libration Width for First-Order Resonances}

For any given first-order mean-motion commensurability (i.e. $q=1$), we wish the estimate the width of the libration domain for circular orbits. In other words, we are interested in the distance $\Delta a^*_{\rm sep}(p)$ between the nominal position of the $(p+1)/p$ resonance and the edge of the inner branch of the separatrix at $e^*=0$.

We begin with the expression (\ref{eq18}) for the Hamiltonian of the SFMR. Since the value of $N^*$ is constant, we can add a quantity equal to $C_1 N^*$ with no change in the dynamics. Thus, we obtain a new expression given by:
\be
F^* = -A_0 {(N^* - S^*)}^2 + (A_1 + C_1) (N^* - S^*) - C_2 \sqrt{2S^*} \cos{\sigma^*} 
\label{eq25}
\ee
or, writing $(N^* - S^*) = L^*/p$,
\be
F^* = -\frac{A_0}{p^2} {L^*}^2 + \frac{(A_1 + C_1)}{p} L^* - C_2 \sqrt{2S^*} \cos{\sigma^*} .
\label{eq26}
\ee
We have kept the denomination of this function, although Hamiltonians (\ref{eq25}) and (\ref{eq26}) differ by a constant amount. We can simplify this expression even further. Using the definition of $L^*_c$ given by (\ref{eq23}) we can write $(A_1+C_1) = 2A_0 L^*_c/p$. Introducing this equality in the Hamiltonian, and after dividing both sides by a factor $A_0/p^2$, we obtain
\be
{\hat F}^* \equiv \frac{p^2}{A_0} F^* = -{L^*}^2 + 2 L^*_c \; L^* - \frac{C_2 p^2}{A_0} \sqrt{2S^*} \cos{\sigma^*} .
\label{eq27}
\ee
To calculate the libration width for a given value of $N^*$ we must first determine the value of the Hamiltonian at the hyperbolic fixed point. From (\ref{eq24}) we can write the value of $S^*_N$ and $L^*_N$ at a given fixed point as:
\be
{\sqrt{2S^*_N}} = -\frac{C_2 p}{2A_0} \biggl( L^*_N - L^*_c \biggr)^{-1} ,
\label{eq28}
\ee
where both $S^*_N$ and $L^*_N$ are function of $N^*$. Substituting into (\ref{eq27}), the Hamiltonian of the hyperbolic fixed point will be given by:
\be
{\hat F}^*_{\rm hyper}(N^*) = -{L^*_N}^2 + 2 L^*_c \; L^*_N + \frac{C_2 p^2}{A_0} \sqrt{2S^*_N}.
\label{eq29}
\ee

We now search for those values of $L^*=L^*_{\rm sep}$ and $S^*=S^*_{\rm sep}$ such that the Hamiltonian at $\sigma^*=0$ attains the same value. This is simply:
\be
{\hat F}^*_{\sigma^*=0}(N^*) = {L^*_{\rm sep}}^2 + 2 L^*_C \; L^*_{\rm sep} - \frac{C_2 p^2}{A_0} \sqrt{2S^*_{\rm sep}}.
\label{eq30}
\ee
Note, however, that $L^*_{\rm sep}$ and $S^*_{\rm sep}$ are not independent, but related through the chosen value of $N^*$. Without any loss of generality, we can then write:
\be
L^*_{\rm sep} = L^*_N + \Delta L^*  \hspace*{0.5cm} ; \hspace*{0.5cm} S^*_{\rm sep} = S^*_N + \Delta S = S^*_N - \frac{1}{p} \Delta L^*, 
\label{eq31}
\ee
where the last equality stems from the constraint that $N^*=const$. Since we are interested in calculating the libration half-width for circular orbits (i.e. $S^*_{\rm sep}=0$), the second equation reduces to $S^*_N = \Delta L^*/p$. Introducing these relations into expressions (\ref{eq29}) and (\ref{eq30}) and equating the values of both Hamiltonians, we obtain
\be
( \Delta L^* )^2 + 2 (L^*_N - L^*_c) \; \Delta L^* + \frac{C_2 p^{3/2}}{A_0} \sqrt{2\Delta L^*} = 0 ,
\label{eq32}
\ee
which admits the non-trivial solution:
\be
 \Delta L^* \equiv L^*_{\rm sep} - L^*_N = p \biggl( -\frac{3 C_2}{4 A_0} \biggr)^{2/3} . 
\label{eq33}
\ee

We can now calculate the distance $\Delta L^*_{\rm sep}$ between the inner separatrix and the exact resonance as
\be
\Delta L^*_{\rm sep} = L^*_{\rm sep} - L^*_{\rm res} \simeq (L^*_{\rm sep} - L^*_N) + (L^*_N - L^*_c) = \biggl( 1 + \frac{\sqrt{2}}{3} \biggr) p \biggl( -\frac{3 C_2}{4 A_0} \biggr)^{2/3} . 
\label{eq34}
\ee
From (\ref{eq12}) and (\ref{eq19}) we can write
\be
-\frac{3 C_2}{4 A_0} =  \frac{m_1}{m_0} \frac{{L^*_c}^{3/2}}{2 p^2} \; \alpha_{\rm res} \; {\hat g}_{1,0}(\alpha^*_{\rm res}) \simeq \frac{0.4}{p} \frac{m_1}{m_0} {L^*_c}^{3/2} , 
\label{eq35}
\ee
where we have used the approximation $\alpha^*_{\rm res} \; {\hat g}_{1,0}(\alpha^*_{\rm res}) \simeq -0.8 p$ (Malhotra 1998). Last of all, converting the results to semimajor axis, we obtain
\be
\Delta a^*_{\rm sep}(p) \simeq 1.6 \; a_1 \; \biggl( \frac{m_1}{m_0} \biggr)^{2/3} \biggl( \frac{m_0}{m_0+m_1} \biggr)^{1/3} \frac{p}{\;\;(p+1)^{2/3}} ,
\label{eq36}
\ee
which is very similar to the expression given by Wisdom (1980). However, in order to preserve accuracy even for low-degree resonances, we will avoid any analytical transformation between $p$ its value of $a^*_{\rm res}$.

\subsection{Separation Between First-Order Resonances}

For this step we will follow the deduction presented in Morbidelli (1999) although, once again, trying to circumvent any non-essential simplifications that may affect accuracy of our model.

The nominal location of the $(p+1)/p$ and $(p+2)/(p+1)$ MMRs can be written as:
\be
a^*_{\rm res} = a_1 \biggl( \frac{m_0}{m_0+m_1} \biggr)^{1/3} \biggl( \frac{p}{p+1} \biggr)^{2/3}
\hspace*{0.3cm} ; \hspace*{0.3cm}
{a^*}'_{\rm res} = a_1 \biggl( \frac{m_0}{m_0+m_1} \biggr)^{1/3} \biggl( \frac{p+1}{p+2} \biggr)^{2/3}
\label{eq37}
\ee
from which their separation is given by:
\be
\Delta a^*_{\rm res}(p) = a_1 \biggl( \frac{m_0}{m_0+m_1} \biggr)^{1/3} \biggl[ \biggl( \frac{p+1}{p+2}  \biggr)^{2/3} - \biggl( \frac{p}{p+1}  \biggr)^{2/3} \biggr] .
\label{eq38}
\ee
We now assume that $p \gg 1$, and expand each of the terms inside the square brackets as a Taylor series. Retaining only first-order terms, we can approximately write:
\be
\begin{split}
\biggl( \frac{p}{p+1} \biggr)^{2/3} &= \biggl( 1 - \frac{1}{p+1} \biggr)^{2/3} \simeq 1 - \frac{2}{3} \frac{1}{(p+1)} \\
\biggl( \frac{p+1}{p+2} \biggr)^{2/3} &= \biggl( 1 - \frac{1}{p+2} \biggr)^{2/3} \simeq 1 - \frac{2}{3} \frac{1}{(p+2)} .
\end{split}
\label{eq39}
\ee

Finally, introducing both expressions into (\ref{eq38}), and after some simple algebra, we obtain:
\be
\Delta a^*_{\rm res}(p) \simeq \frac{2}{3} a_1 \biggl( \frac{m_0}{m_0+m_1} \biggr)^{1/3} \frac{1}{(p+1)(p+2)} .
\label{eq40}
\ee
This is moderately different from the expression in Wisdom (1980) and Deck et al. (2013) which adopt $\Delta a^*_{\rm res} = (2 a_1) / (3(p+1)^2)$. On one hand, it is important to explicitly keep the mass ratio between the perturber and the star, which might be important for massive planets, or even binary stellar systems. On the other hand, the difference in the dependence on $p$ is also significant, especially for low-degree resonances. It is simple to see that while the approximate relation adopted by Wisdom (1980) systematically overestimates the true separation, the opposite occurs for (\ref{eq40}). We found that a much more accurate estimate, even for low values of $p$ may be written as:
\be
\Delta a^*_{\rm res} (p) \simeq \frac{2}{3} a_1 \biggl( \frac{m_0}{m_0+m_1} \biggr)^{1/3} \frac{1}{(p+1)(p+3/2)} ,
\label{eq41}
\ee
in other words, half way between both of the original approximations.

\subsection{The Classical Overlap Condition}

Having expressions for the libration half-width (eq. (\ref{eq36})) and the separation between consecutive first-order resonances (eq. (\ref{eq41})), we can now proceed to calculate the condition for resonance overlap. Following Wisdom (1980), this is said to occur whenever the order of the resonance acquires a value $p = p_c$ such that:
\be
\Delta a^*_{\rm res}(p_c) = 2 \Delta a^*_{\rm sep}(p_c) .
\label{eq42}
\ee
Deck et al. (2013) improved this estimation replacing the right-hand side with $\Delta a^*_{\rm sep}(p_c) + \Delta a^*_{\rm sep}(p_c+1)$, leading to a significant change in the results.
Nevertheless, as will be discussed further on, this modification of the classical condition will not prove necessary and we will adopt expression (\ref{eq42}).

Whatever our choice, the idea remains the same: overlap is defined when the inner separatrix of the $(p_c+1)/p_c$ MMR intersects the outer branch of the $(p_c+2)/(p_c+1)$ resonance. As in the original Chirikov criteria, the crossing of separatrices of two commensurabilities generates a dynamical route leading to orbital instability. However, it is important the stress that mere proximity between resonances is not sufficient; overlap requires the existence and intersection of two separatrix. If both are not present, then overlap cannot be said to occur.

Solving equation (\ref{eq42}), we obtain the value of $p_c$ that signals resonance overlap (for circular orbits) according to this model as:
\be
4.8 \; p_c (p_c+3/2)(p_c+1)^{1/3} \simeq \biggl( \frac{m_1}{m_0} \biggr)^{-2/3}.
\label{eq43}
\ee
Since giving results in values of $p_c$ is awkward, we can transform them to values of $a^*$ using expression (\ref{eq37}). As before, we avoid analytical simplifications and any a-priori assumption that $p_c \gg 1$, but choose to perform a one-dimensional numerical fit of the critical semimajor axis as function of $m_1/m_0$. The result yields:
\be
a^*_{\rm overlap} \simeq a_1 \biggl[ 1 - 1.225 \; \biggl( \frac{m_1}{m_0} \biggr)^{0.28} \biggr] .
\label{eq44}
\ee
This, then, is the minimum initial mean semimajor axis, for circular orbits, such that the massless particle lies in an unstable region generated by the overlap of first-order interior MMRs with a perturbing planet (also in circular orbit) of mass $m_1$. 

It is important to stress that this calculation, as well as the resulting expression, is given in {\it averaged} orbital elements. This is vital when attempting to compare its predictions with numerical simulations of the exact Newtonian differential equations. To obtain the overlap limit in osculating semimajor axis we must perform the back transformation, which yields the following approximate relation
\be
a_{\rm overlap} \simeq a_1 \biggl[ 1 - 1.06 \; \biggl( \frac{m_1}{m_0} \biggr)^{0.275} \biggr] .
\label{eq45}
\ee
This expression can now be compared directly with numerical integrations, as long as the initial angles are chosen equal to zero. Note that the exponents of both descriptions of the overlap criterion are similar to $2/7 \simeq 0.286$, but slightly smaller.

\section{Numerical Tests of the Overlap Criteria}

\subsection{The $\Delta e$ Indicator}

An interesting tool to analyze the behavior of planetary systems is the so-called Maximum Eccentricity Method (MEM) (e.g. Dvorak et al. 2004), which follows the maximum value of $e$ attained by a given body during a numerical simulation. For initially eccentric orbits, a better indicator is the difference between the maximum and minimum values, or the value of $\Delta e$. 

Although it is not a measure of chaotic motion, this $\Delta e$ indicator is an extremely useful tool to map the resonant structure in N-body problems. Its application is very simple. A grid of initial conditions in a representative plane is integrated numerically for a timespan $T$ larger than the period of the slowest angle of the system. During the integration we keep track of the minimum and maximum values attained by one of the actions (call it $J$), and calculate the difference $\Delta J = J_{\rm max} - J_{\rm min}$. Finally, we plot the value of $\Delta J$ as a color graph in the plane of initial conditions. 

\begin{figure}[t!]
\centering
\includegraphics*[trim=0cm 1cm 0cm 5.5cm,clip=true,width=12cm]{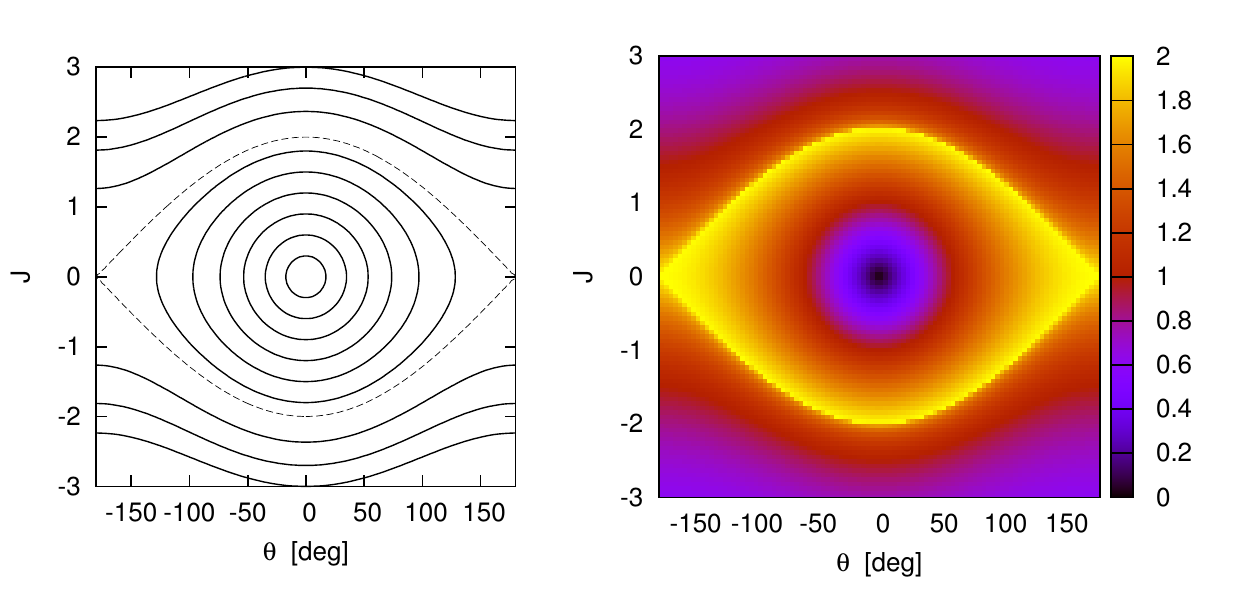}
\caption{{\bf Left:} Phase plane of the simple pendulum. {\bf Right:} Maximum increment of the momentum $J$ for a grid of initial conditions in the plane.}
\label{fig3}
\end{figure}

Figure \ref{fig3} shows an example for the simple pendulum. The left-hand frame shows the level curves of the Hamiltonian in the $(J,\theta)$ plane. Since the system has only a single degree of freedom, the plane is the complete phase space. The stable fixed point appears at the center of the graph ($J=0$, $\theta=0$), while the separatrix is highlighted in dashed lines. 

We now divide the plane in a $100 \times 100$ grid of initial values of the action and the angle. Each initial condition is then integrated for a total time $T$. The resulting values of $J$ are shown in a color plot on the right-hand side of the figure. Initial conditions close to the stable fixed point appear with a dark color, indicating very small values of $\Delta J$. Conversely, initial conditions close to the separatrix appear as very light colored regions, corresponding to large changes in the momentum. Thus, even if we were not able to obtain explicitly the phase curves of the Hamiltonian (i.e. plot on the left), the color plot on the right would allow us to estimate both the location of possible fixed points as well as the separatrix of the resonance region. 

\begin{figure}[t!]
\centering
\includegraphics[trim=0cm 0.2cm 0cm 0.2cm,clip=true,width=13cm]{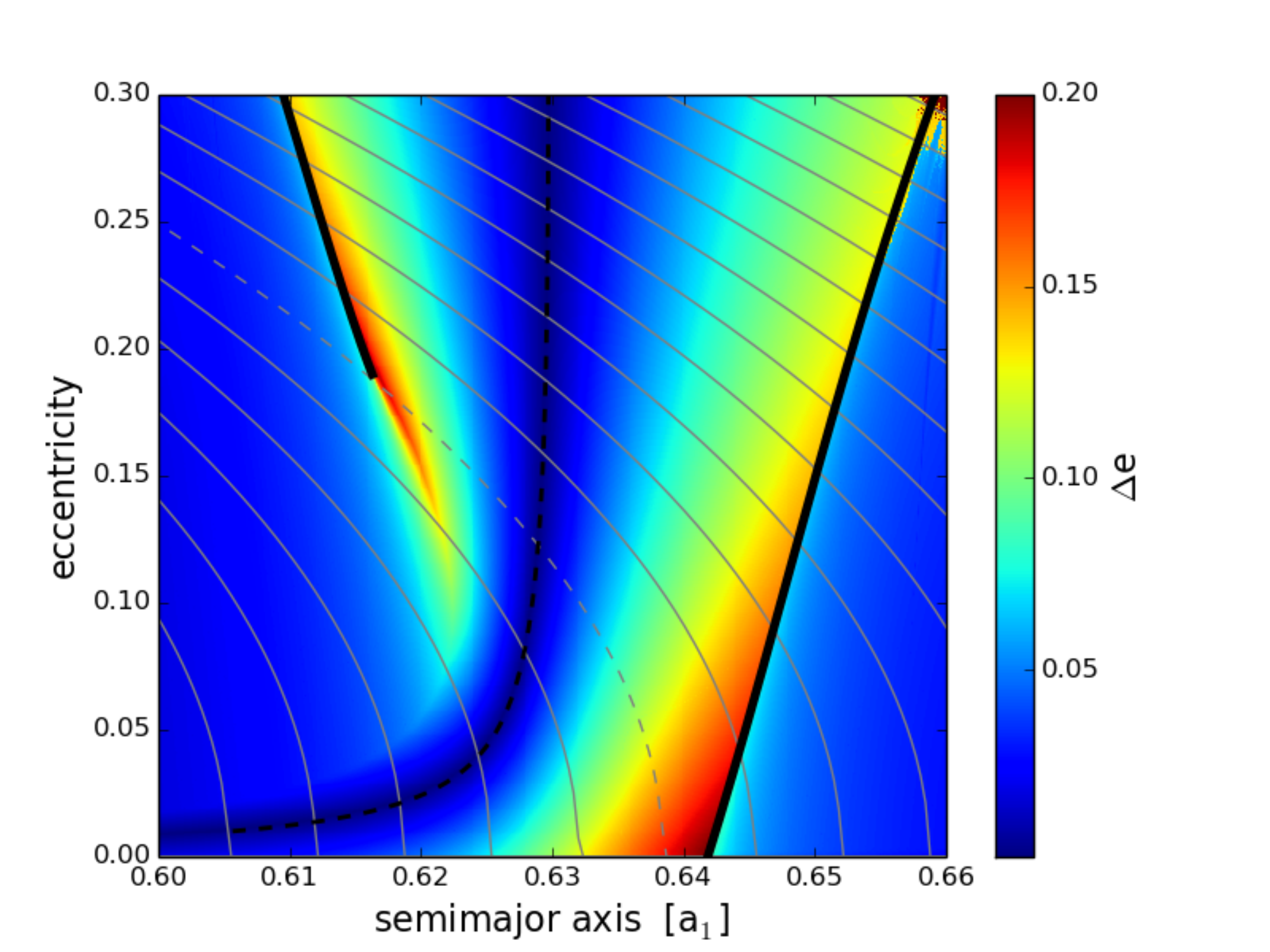} 
\caption{Values of $\Delta e$ in the representative plane $(a/a_1,e)$ of osculating initial conditions in the vicinity of the 2/1 MMR with Jupiter (current mass). All angular variables were initially chosen equal to zero. The pericentric branch of zero-amplitude solutions is shown as a dashed black curve and both branches of the separatrix in broad black continuous lines. Curves of constant $N^*$ are indicated with gray lines, with $N_c$ highlighted as a dashed curve. The regions of maximum variation of the eccentricity appear red, while those associated with small changes are indicated in blue.}
\label{fig4}
\end{figure}

Having defined the method, we can apply it to a mean-motion resonance in the circular restricted three-body problem. Figure \ref{fig4} shows the resulting map for the 2/1 MMR in the representative plane $(a/a_1,e)$ of (osculating) initial conditions, where all angles were chosen initially equal to zero. All initial conditions were integrated with a Bulirsch-Stoer based N-body code for $T=10^5$ orbits of the perturber, and the color code is the difference in eccentricity attained by each particle (i.e. $\Delta e$). Thin gray lines mark different values of $N^*$, with $N_c$ highlighted with a dashed curve.

We can now compare these results to the predictions of the semi-analytical resonance model. The predicted pericentric branch is shown as a dashed black curve, while both separatrix branches are indicated with broad black lines. The agreement is very good, indicating that the semi-analytical model is a very precise tool for estimating the features of a given mean-motion resonance. More importantly, it shows that, for circular orbits, there is no outer separatrix for the 2/1 MMRs, and the same holds for any other first-order commensurability. For $e=0$, all initial conditions with $a < a_{\rm res}$ show practically no change in the eccentricity and remain close to circular orbits throughout the integration. Only those initial conditions with $a > a_{\rm res}$ show an increase in the eccentricity, reaching a maximum at the location of the inner branch of the separatrix, roughly located at $a \simeq 0.64 a_1$. 

\begin{figure}[t!]
\centering
\includegraphics[width=14cm]{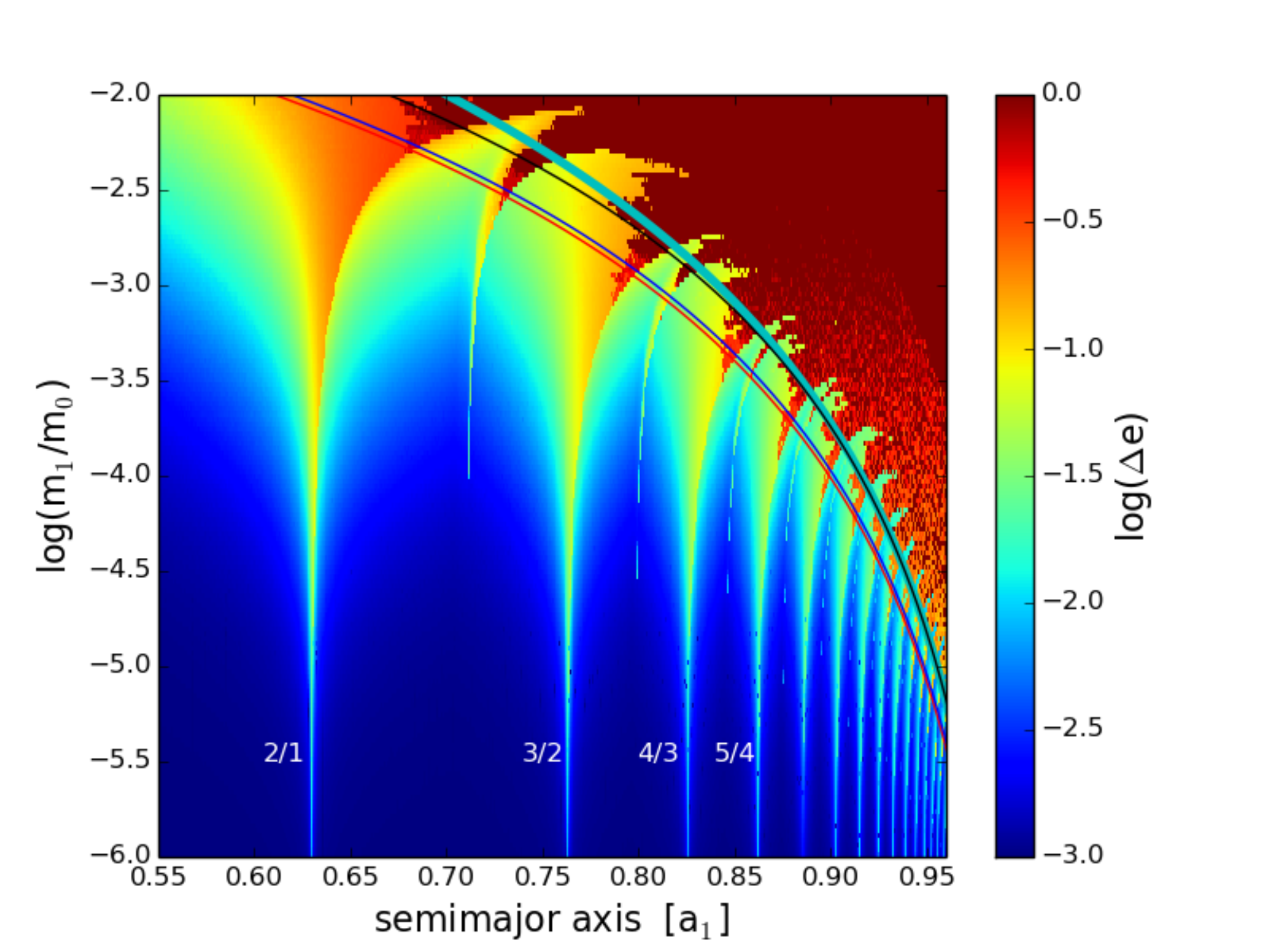}   
\caption{$\Delta e$ map of the $(a/a_1,m_1/m_0)$ plane of initial conditions (in osculating elements) with circular orbits. Final hyperbolic orbits are shown in dark red. The colored curves are different predictions of resonance overlap: Wisdom's criterion with $D = 1.30$ (black), the analytical expression by Deck et al. (2013) with $D=1.46$ (blue), numerical fit by Duncan et al. (1989) with $D = 1.49$ (red), and the criterion developed here (cyan). The location of the main first-order MMRs are indicated with white text.}
\label{fig5}
\end{figure}

Again, the nonexistence of the outer separatrix for low eccentricities is the result of plotting the representative plane in non-canonical variables. For all values of $N^* \ge N_c$ both branches of the separatrix exist, although with different values of the semimajor axis and eccentricity.

\subsection{Large-scale Dynamical Map of the Representative Plane}

In order to compare the different versions of the resonance overlap criterion, we constructed a dynamical map in the plane $(a/a_1,m_1/m_0)$, with 800 values of the osculating semimajor axis and 200 values of the mass ratio between $m_1/m_0 \in [10^{-6},10^{-2}]$. Initial osculating eccentricities and angles where chosen equal to zero, and all orbits were integrated for $T=10^5$ orbital periods of the perturber. Results are shown in Figure \ref{fig5}, where blue regions mark small changes in the eccentricity, and dark red indicate hyperbolic motion and/or escapes.

The map shows the different resonances present in this interval of semimajor axis, starting with the 2/1 MMR at $a/a_1 \simeq 0.63$, up to first-order resonances with $p \rightarrow 20$ for semimajor axis ratios above $0.95$. These resonances generate a ``saw-tooth'' shape limit for the unstable region. As we will see further on, not all correspond to first-order commensurabilities. The same figure shows, in continuous curves, the predictions of the different overlap criteria discussed in the previous section. To allow for an adequate comparison, all the values of $a^*_{\rm overlap}$ were transformed to osculating semimajor axes. 

While all the criteria appear similar, the expressions deduced by Duncan et al. (1989) and Deck et al. (2013) seem a good fit to the lower extrema of the saw-tooth structure. In the case of the value given by Duncan et al. (1989), this is not unexpected since it was obtained by a least-square fit of a series of numerical simulations using a symplectic map. The black curve, corresponding to the original prediction by Wisdom (1980), passes through the middle of the saw-tooth region, as though these structures were evened out. The value $D=1.30$ then appears to yield a better ``average'' boundary between the stable and unstable domains. Last of all, the broad cyan curve presents the predictions of our model, as given by eq. (\ref{eq45}). Although it has a different functional form, it yields practically the same results as Wisdom (1980) for, say, $m_1/m_0 < 10^{-4}$. For larger masses, however, there is an increasing and noticeable discrepancy, and our values are systematically closer to the perturber. 

Before pursuing a more detailed comparison between these models, it is important to review whether the assumptions behind the resonance overlap criterion are in fact consistent with the structures shown in the dynamical maps. First, there is evidence of second-order resonances (5/3, 7/5, 9/7, etc.), particularly close to the instability limit, that interact with the first-order MMRs and contribute to generate the chaotic domain. These are not considered in the overlap criterion developed so far. 

A second feature that can be perceived from the dynamical map is that the outer separatrix of the first-order MMRs appears to have little effect in the dynamics of the system. The reason behind this was mentioned before: the outer separatrix (located at $a < a_{\rm res}$) only exists for eccentricities above a certain threshold, and is not present for circular orbits. Then, the only truly resonant region for circular orbits occurs for $a > a_{\rm res}$ and is characterized by the inner separatrix. Thus, the idea of using the libration width for the inner separatrix as a proxy for the outer branch is questionable, at least for circular orbits.

\begin{figure}[t!]
\centering
\includegraphics[width=14cm]{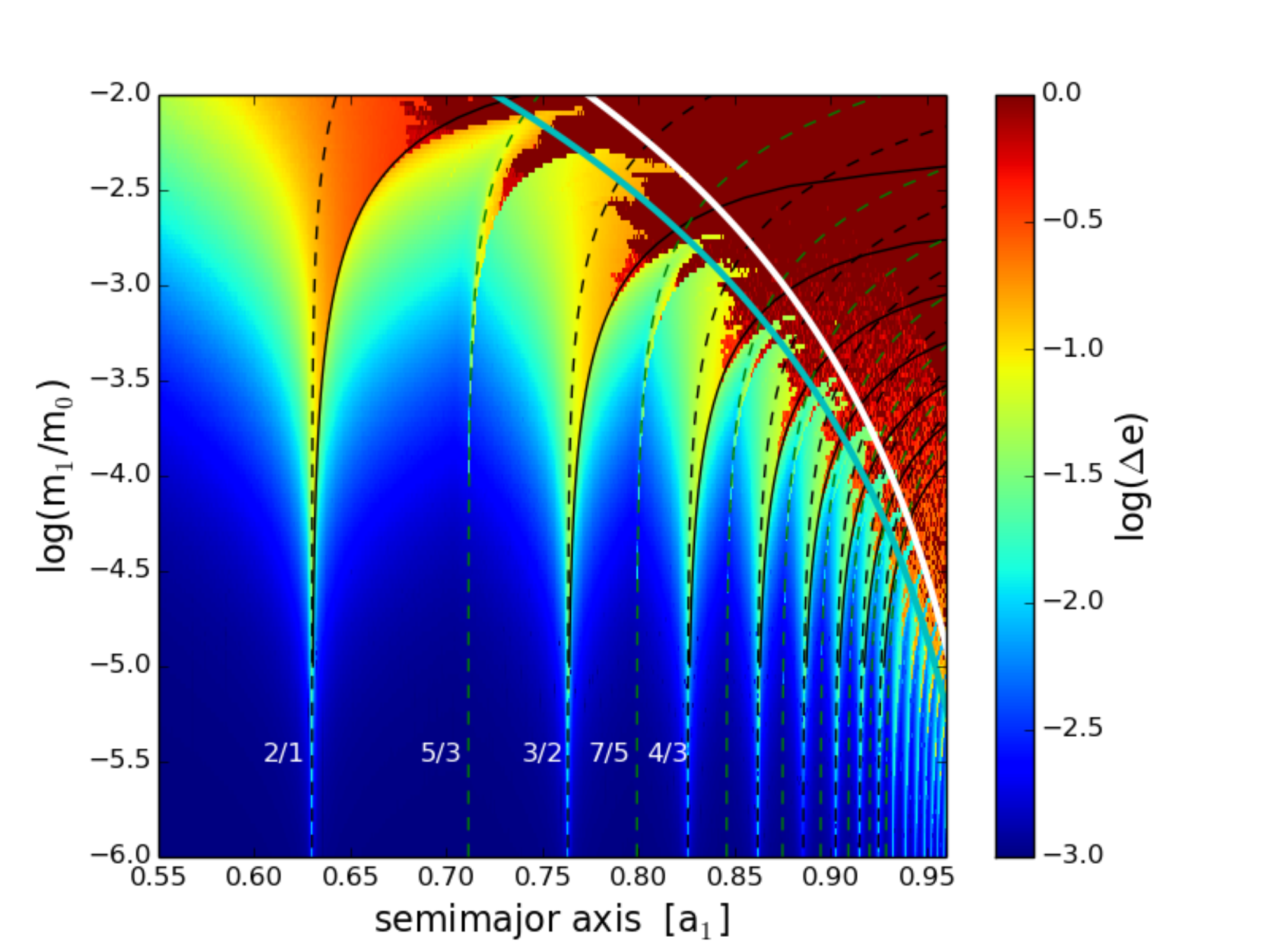} 
\vspace{-3mm}
\caption{Same dynamical map shown in the previous figures, this time superposed with the resonant structure of first (black) and second-order (green) MMRs. The location of each pericentric branch in shown in dashed lines, while the extension of the inner separatrix for circular orbits (first-order resonances only) is shown in continuous lines. The Cyan curve shows the result of the new resonance overlap criterion developed using these resonances. The broad white curve, corresponding to eq. (\ref{eq49}), is the approximate location of the beginning of the global unstable region.}
\label{fig6}
\end{figure}

\begin{figure}[t!]
\centering
\includegraphics[width=12cm]{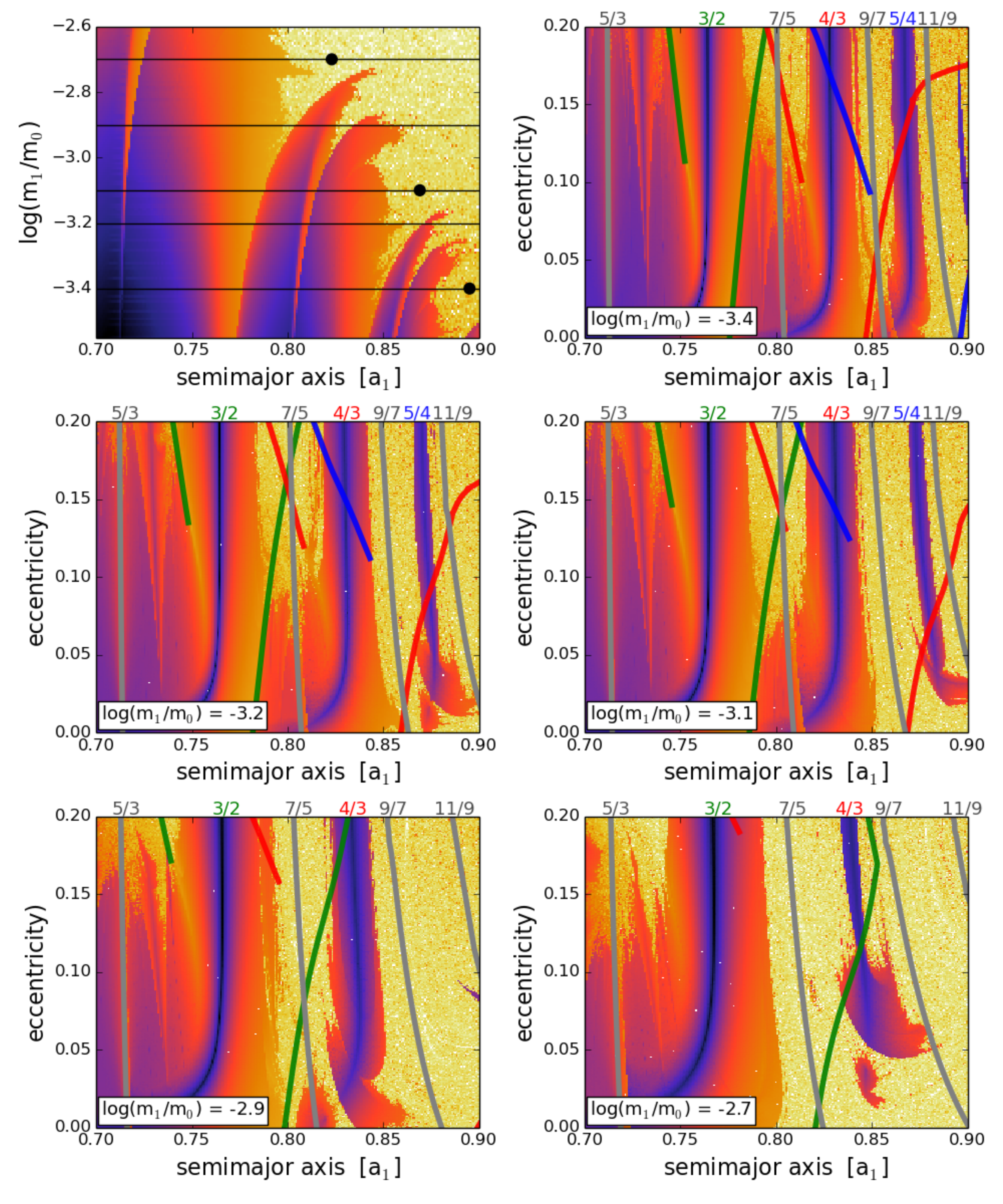} 
\vspace{-3mm}
\caption{Top left-hand frame shows a zoom of the dynamical map in the $(a/a_1,m_1/m_0)$ plane presented in the previous figure. Five values of the perturbing mass are highlighted with horizontal black lines. The intersection between adjacent first and second-order resonances are indicated with black circles. The remaining graphs present new dynamical maps in the $(a/a_1,e)$ plane for each value of $m_1/m_0$. The location of the second-order resonances are indicated with gray lines, while the separatrix of first-order MMRs are shown in different colors. Each commensurability is indicated on the top of the frames.}
\label{fig7}
\end{figure}

\subsection{New Criterion with 1$^{\rm st}$ and 2$^{\rm nd}$-order MMRs}

To analyze this point in more detail, Figure \ref{fig6} shows the same dynamical map as before, only this time we have superimposed the nominal position of the first and second-order MMRs (dashed black and green lines, respectively) in terms of the osculating semimajor axis. While $a^*_{\rm res}$ is a weak function of the perturbing mass (and actually decreases for larger values of $m_1$), the difference between $a^*$ and $a$ shows a much stronger dependence. Moreover, our choice of angles for the representative plane (i.e. $\lambda=\lambda_1=\varpi=0$) corresponds to the maximum value attained by the semimajor axis due to short-period oscillations, which leads to exact resonance occurring closer to the planet for increasing $m_1$. 

The same figure also shows, in continuous black curves, the inner separatrix for the first-order commensurabilities. We can now see a much clearer agreement between the resonant structure and the results of the numerical simulations. As we consider larger values of $m_1$, the region close to the inner separatrix of each first-order resonance becomes increasingly chaotic, generating a region of orbital instability, characterized by $\Delta e \rightarrow 1$. Concurrently, a different chaotic domain appears, linked to  second-order commensurabilities, whose effects have been almost negligible until that point. The size of both chaotic regions increase with the perturbing mass until, at some point, both intersect. Global chaos then sets in and most initial conditions between 
both commensurabilities become unstable. 

A complementary view is presented in Figure \ref{fig7}, in the form of new maps in the $(a/a_1,e)$ plane for five values of $m_1/m_0$. As before, dark regions are associated to small changes in the eccentricity during the total integration time, while the opposite occurs for initial conditions identified with lighter tones. The location of second-order resonances are indicated in gray lines, while the separatrix of first-order commensurabilities are shown in color curves. These were calculated using a semi-analytical model (e.g. Beaug\'e 1994) and not with the SFMR to allow for a better correlation with the numerical results. It is important to keep in mind, however, that the separatrix have been drawn assuming isolated resonances and, as such, do not take into account perturbations form nearby commensurabilities.

While the dynamical features are complex and show evidence of the interaction between many different resonances, some characteristics may be deciphered analyzing their evolution as function of the planetary mass. For $m_1/m_0 = 10^{-3.4}$, all initially circular orbits with $a/a_1 > 0.88$ are unstable, a value close to the intersection point between the 11/9 resonance and the inner separatrix of the 5/4 commensurability. Initial conditions with smaller semimajor axis appear primarily regular, although some chaotic motion is visible in the vicinity of $a/a_1 \simeq 0.85$, associated to a near-intersection between the 9/7 resonance and the inner separatrix of the 4/3. 

Between $m_1/m_0 = 10^{-3.2}$ and $m_1/m_0 = 10^{-3.1}$, the 9/7 and 4/3 MMRs intersect and all circular initial conditions closer to the planet are unstable. A new isolated region of chaotic motion appears at $a/a_1 \simeq 0.80$, generated by the interaction between the 7/5 commensurability and the inner separatrix of the 3/2 resonance. For $m_1/m_0 = 10^{-2.9}$ this chaotic layer becomes more extensive and finally merges with the global instability region for 
$m_1/m_0 \simeq 10^{-2.7}$.

The same behavior can be observed for other values of the planetary mass and semimajor axis. Thus, it appear that global instability may be estimated by the intersection of the inner separatrix (i.e. located at $a > a_{\rm res}$) of a given first-order MMR with the nominal location of its adjacent second-order commensurability (i.e. black circles in the top left-hand frame of Figure \ref{fig7}). In fact, the libration width of the second-order resonance does not appear important at all. This result enormously simplifies the calculations, since we do not require to model the resonance structure of the second-order resonances; all we require is to estimate its location in the semimajor axis domain for any given planetary mass.

These numerical results led us to postulate a new overlap criterion based in the interaction between adjacent first and second-order MMRs. To calculate this new criterion, we first write average mean motion of a generic first-order MMR and its adjacent second-order commensurability as:
\be
n^*_{\rm res} = \frac{p+1}{p} n_1  \hspace*{0.5cm} ; \hspace*{0.5cm} {n^*}'_{\rm res} = \frac{2p+3}{2p+1} n_1.
\label{eq46}
\ee
This expression excludes ``false'' second-order resonances such as the 4/2 or 6/4, since both the numerator and denominator present in ${n^*}'_{\rm res}$ are odd numbers. The distance between them can be calculated using the same procedure as in the Section 3.2, which now gives:
\be
\Delta a^*_{\rm res}(p_c) = a_1 \biggl( \frac{m_0}{m_0+m_1} \biggr)^{1/3} \biggl[ \biggl( \frac{2p_c+1}{2p_c+3}  \biggr)^{2/3} - \biggl( \frac{p_c}{p_c+1}  \biggr)^{2/3} \biggr] .
\label{eq47}
\ee
For the libration half-width of the first-order commensurabilities we use expression (\ref{eq36}), analogous to our derivation of the classical overlap criterion. Equating (\ref{eq47}) with (\ref{eq36}) we can solve for $p_c$ such that $\Delta a^*(p_c) = \Delta a^*_{\rm res}(p_c)$. After transforming the result in terms of the osculating semimajor axis $a$, we finally obtain our new overlap condition as:
\be
a_{\rm overlap} \simeq a_1 \biggl[ 1 - 0.91 \biggl( \frac{m_1}{m_0} \biggr)^{0.26} \biggr] ,
\label{eq48}
\ee
where both numerical coefficient were determined from a non-linear regression with values of $m_1/m_0 \in [10^{-7},10^{-1}]$. Figure \ref{fig6} shows, with a broad Cyan curve, the prediction of this new criterion. Since the calculations are not exact, there is a slight difference with respect to the actual resonance intersections. It is interesting to note that this result is very similar to the one determined using the classical overlap condition (i.e. Figure \ref{fig5}), although both were obtained with completely different assumptions. 

Since the SFMR underestimates the separatrix width (see Figure \ref{fig2}), the predictions of expression (\ref{eq48}) fall short of the instability limit as shown in Figure \ref{fig7}. However, we found that it is possible to improve the estimation simply changing the numerical factor. The white curve in Figure \ref{fig6} corresponds to the equation
\be
a_{\rm unstable} \simeq a_1 \biggl[ 1 - 0.75 \biggl( \frac{m_1}{m_0} \biggr)^{0.26} \biggr] ,
\label{eq49}
\ee
which has the same functional dependence with the perturbing mass as the new overlap criterion, but with a smaller coefficient. Although this is an empirical correction of our analytical result, it results in a fairly accurate estimate for the value for the osculating semimajor axis that marks the beginning of the region of global instability.

\section{The Hill Stability Criterion}

A different stability criterion may be defined in terms of the value of the Jacobi constant $C_J$ of the particle, as compared with its value $C_{L_1}$ at the $L_1$ Lagrange point. If $C_J > C_{L_1}$, then the massless body is forever trapped in a zero-velocity curve that encompasses the central mass $m_0$, but which does not contain the perturbing mass $m_1$. In such a case the initial condition is said to be Hill Stable. Conversely, if the initial conditions are such that $C_J < C_{L_1}$, then the stability is not guaranteed, and may suffer close approaches with $m_1$ and become temporarily trapped by the smaller primary. Its $m_0$-centric motion will then be characterized by hyperbolic orbits and thereby cataloged as unstable. 

The Jacobi constant then constitutes a valuable asset with which to determine the (Hill) stability of a given massless particle in the realm of the CR3BP. We will make use of this feature to develop a complementary stability criterion that will later be compared with the predictions of the resonance overlap criterion.

\subsection{Hill Stability in Orbital Elements}

The main problem with the practical application of this criterion is to calculate both the Jacobi constant for $L_1$ and its value for any given initial condition in orbital elements. Equations usually found in the literature either make use of series expansions (which may or not yield accurate results for large planetary masses) or require Cartesian coordinates and velocities in the rotating reference frame. In the next paragraphs we will introduce a way to overcome both limitations.

Seidov (2004) introduced closed formulas to calculate the position and Jacobi constant for $L_1$. Let us call $r_{L_1}$ the distance from $m_0$ to $L_1$. We can then write
\be
r_{L_1} = a_1 (1 - \delta_{L_1})
\label{eq50}
\ee
where the new auxiliary quantity will satisfy $\delta_{L_1} \rightarrow 0$ as $m_1 \rightarrow 0$.
Seidov (2004) found that $\delta_{L_1}$ and the planetary mass are related through:
\be
\frac{m_1}{m_0} = \frac{(1 - \delta_{L_1})^3 (1 + \delta_{L_1} + \delta_{L_1}^2)}{\delta_{L_1}^3 (3 - 3\delta_{L_1} + \delta_{L_1}^2)} .
\label{eq51}
\ee
Although this equation is exact for any value of $m_1$, it is implicit in $\delta_{L_1}$. Therefore, in order to determine the position of $L_1$ as function of the planetary mass, we must solve it using successive approximations. Since $\delta_{L_1}$ is usually a small quantity, we can rewrite (\ref{eq51}) is a form more adequate for the iterative process. This is:
\be
\delta_{L_1}^3 = \frac{1}{3} (1 - \delta_{L_1})^3 (1 + \delta_{L_1} + \delta_{L_1}^2)
                  \biggl( \frac{m_1}{m_0} \biggr) + \delta_{L_1}^4 - \frac{1}{3} \delta_{L_1}^5 .
\label{eq52}
\ee
Choosing zero as the initial guess in the right-hand side, this expression can be solved in just a few iterations, and yields very precise results even for large values of the perturbing mass. 

Having determined the location of the Lagrange point, the value of the Jacobi constant $C_{L_1}$ can also be calculated using a closed formula. Again following Seidov (2004), we can write:
\be
C_{L_1} = \frac{{\cal G}(m_0+m_1)}{a_1} \biggl(\frac{3 - 12 \epsilon + 15 \epsilon^2 - 10 \epsilon^3 - 4 \epsilon^4}{(1-2\epsilon-\epsilon^2)^2} \biggr) \hspace*{0.5cm} ,
\label{eq53}
\ee
where $\epsilon = \delta_{L_1} ( 1 - \delta_{L_1} )$.
 
Our next task is the obtain an explicit equation for $C_J$ for any massless particle in terms of its orbital elements. Although the Jacobi constant is usually expressed in Cartesian coordinates in the rotating reference frame, it is also possible to write it in an inertial frame (e.g. Murray and Dermott 1999). Assuming initial conditions $\lambda = \lambda_1 = \varpi = 0$ (i.e. the particle is located in its pericenter and all bodies are aligned), then we find:
\be
C_J = 2 n_1 \biggl( {\cal G} m_0 a (1-e^2) \biggr)^{1/2} + 2 {\cal G} \biggl( \frac{m_0}{\Delta_0} + \frac{m_1}{\Delta_1} \biggr) - \frac{{\cal G} m_0}{a} \biggl( \frac{1+e}{1-e} \biggr) ,
\label{eq54}
\ee
where $n_1$ is the mean motion of the perturbing planet, and $\Delta_i$ are the distances from the massless body to $m_i$, given by:
\be
\begin{split}
\Delta_0 &= a(1-e)   \\
\Delta_1 &= a_1 - a(1-e) .
\end{split}
\label{eq55}
\ee

With these simple expressions we may now calculate the boundary of the Hill stability limit. Given values of $m_0$, $m_1$ and $e$, we search for the value of the semimajor axis $a_{\rm Hill}$ such that $C_J = C_{L_1}$. This procedure can be performed for a range of planetary masses $m_1$, from which we can construct our stability limit curve in the $(a/a_1,m_1/m_0)$ plane. 

An important point is that, contrary to the resonance overlap criterion, all the expressions developed here are given in terms of osculating orbital elements. The only drawback is that this
criterion cannot be expressed in a simple closed form for arbitrary values of $m_1$. Although some approximations may be found in the literature (e.g. Gladman 1993, Deck et al. 2013), usually based on the Hill description of the restricted three-body problem, they are usually only accurate for small perturbing masses (of the order of $m_1/m_0 \sim 10^{-4}$ and lower), and even then just in the circular case. In comparison, the model developed above is still purely analytical and extremely fast to use once implemented in a computer code. 

\begin{figure}[t!]
\centering
\includegraphics[width=14cm]{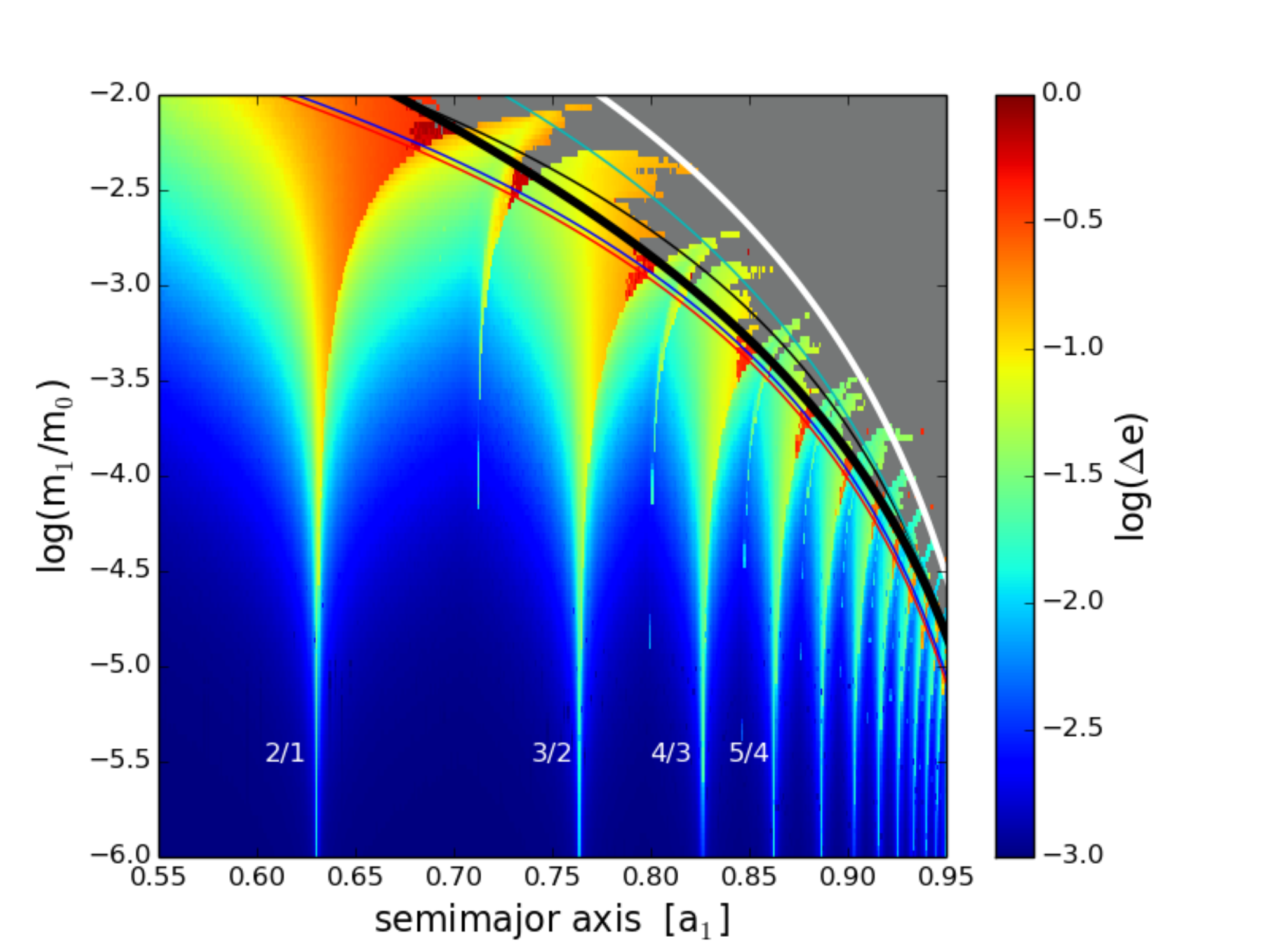}  
\vspace{-3mm}
\caption{Dynamical map for initially circular orbits (i.e. $e=0$) and all angular variables equal to zero, for a grid of initial conditions in the $(a/a_1,m_1/m_0)$ plane. This map is equivalent to the one discussed extensively in the previous section, except that all unstable initial conditions are highlighted in gray. The thin curves mark the predictions of the different resonance overlap criteria; Black: Wisdom (1980), Blue: Deck et al. (2013) and Red: Duncan et al. (1989), Cyan: new overlap limit defined in section 4.3. The broad black curve shows the predictions of the Hill Stability Criterion, while the broad white curve is the empirical global instability limit $a_{\rm unstable}(m_1/m_0)$ given by eq. (\ref{eq49}).}
\label{fig8}
\end{figure}

\begin{figure}[t!]
\centering
\includegraphics[width=14cm]{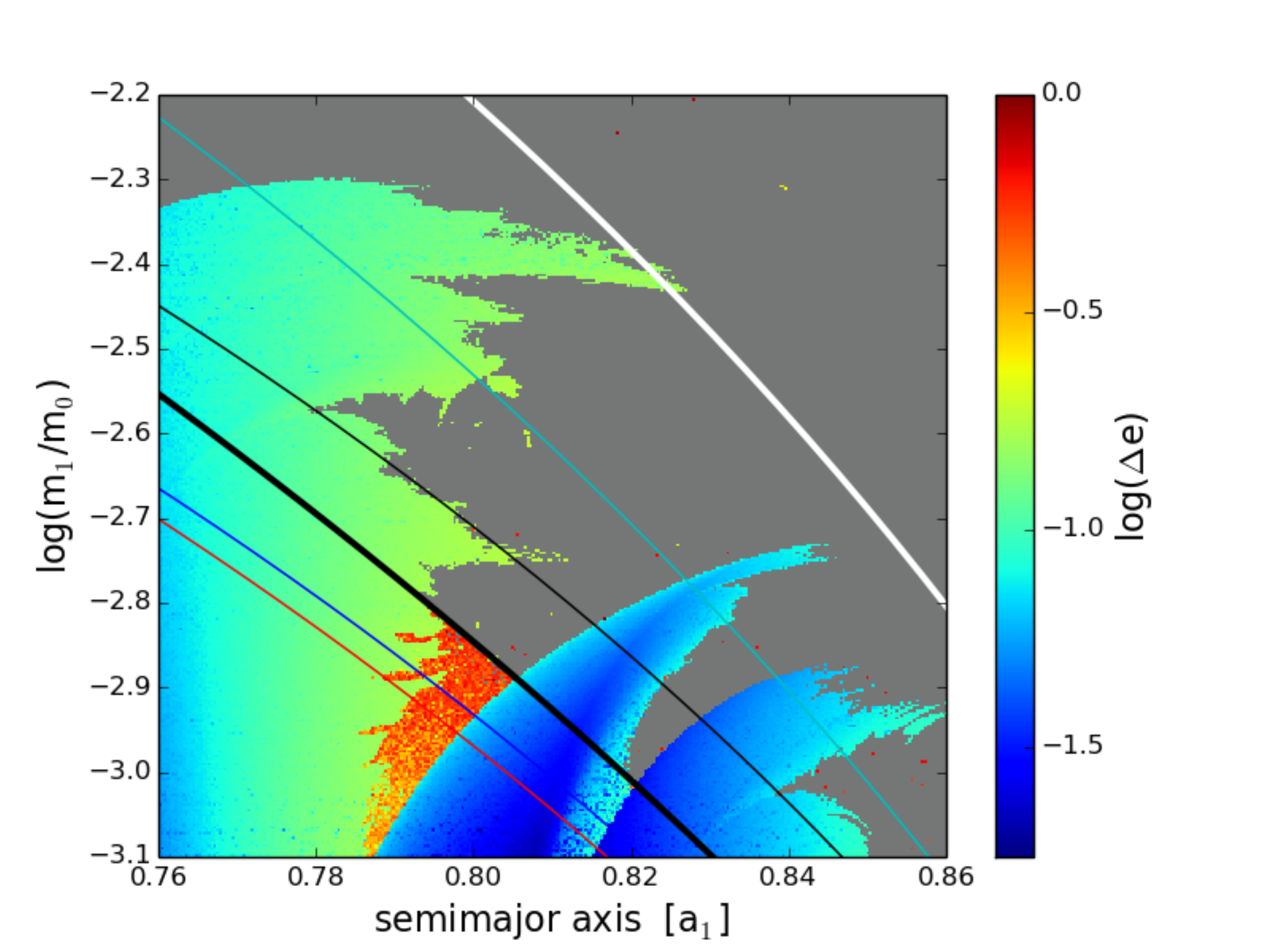} 
\vspace{-3mm}
\caption{Same as previous Figure, but now zooming in on a spike centered roughly around the 3/2 mean-motion resonance. Notice the change in stability on both sides of the Hill limit.}
\label{fig9}
\end{figure}

\subsection{Application to Circular Orbits}

In order to compare the predictions of this criterion with those obtained from resonance overlap, we will first analyze the case $e=0$. Results are shown in Figure \ref{fig8}, again in the foreground of the same dynamical map discussed in the previous section. However, we have also highlighted in gray all initial conditions that are unstable within the integration timespan. These include cases where the final orbit is hyperbolic as well as those that cross the orbit of the perturber while retaining $e < 1$. The broad white curve is the empirical proxy for $a_{\rm unstable}$, while the broad black curve is the prediction of the present Hill Stability Criterion $a_{\rm Hill}$. All versions of the resonance overlap criterion are shown in thin continuous lines.

Figure \ref{fig9} shows a zoom of the region around the spike near the 3/2 MMR. This plot was constructed from a fresh set of numerical simulations covering a 300$\times$300 grid in semimajor axis and mass ratios. As before, all angles were initially set equal to zero and the integration time for each initial condition was $T=10^5$ orbits of the perturber. Each spike shows a complex shape on the left side, but a significantly more smooth edge on the right side. The origin of the dichotomy is not clear, although it could be caused by higher-order resonances that are only visible in this level of detail. 

As always, thin continuous lines shows the result of the different resonance overlap criteria, while the broad white curve is the proxy for the beginning of the global instability region. Except for a big wedge around $m_1/m_0 \simeq 10^{-2.4}$, this prediction of the beginning of the chaotic sea appears very accurate. Again, with the exception of the wedge, all initial conditions with $a > a_{\rm unstable}$ (i.e. beyond the broad white curve) are unstable. 

The broad black curve shows the results of the Hill Stability limit. Although the saw-tooth structure seems to be indifferent to the presence of the limit curve, the values of $\Delta e$ are different on both sides. All orbits satisfying the condition $a > a_{\rm Hill}$ are in fact unstable, lending credibility to our expressions in terms of the orbital elements. For smaller semimajor axes, however, the increase in eccentricity is significantly lower, indicating that, although there is a certain excitation of the orbit, it is bounded and not sufficient to render the motion unstable. 

It is interesting to note that the region located between both stability criteria (i.e. $a_{\rm Hill} < a < a_{\rm unstable}$) is also characterized by sharp borders between stable and unstable orbits. More importantly, the condition $C_J > C_{L_1}$ is sufficient but not necessary for stability; initial conditions exist for which motion is always bounded to the central mass even though the zero-velocity surface includes both primaries. 

\begin{figure}[t!]
\centering
\includegraphics[width=12cm]{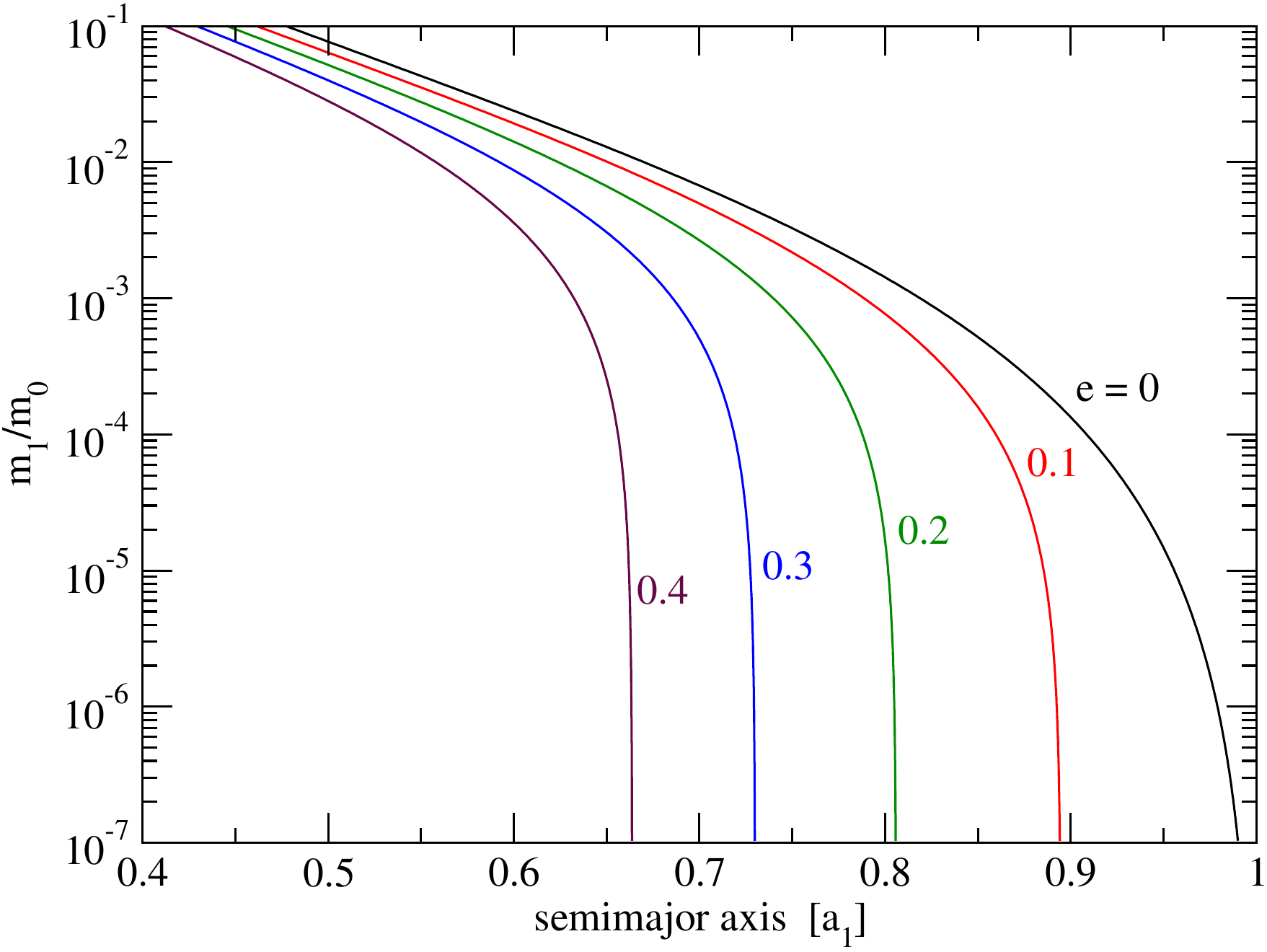}
\vspace{-3mm}
\caption{The Hill Stability limit for different values of the particle's eccentricity (identified by the number accompanying each curve).}
\label{fig10}
\end{figure}

\subsection{Application to Eccentric Orbits}

Figure \ref{fig10} shows the results of the calculation of the Hill Stability limit, in the $(a/a_1,m_1/m_0)$ plane, for different values of $e$. The boundary for low masses is very sensitive to the initial eccentricity, while the change seems less drastic for larger masses. Even so, even a low eccentricity introduces a significant change in the stability limit, which should be noticeable in a dynamical map. 

\begin{figure}[t!]
\centering
\includegraphics[width=13cm]{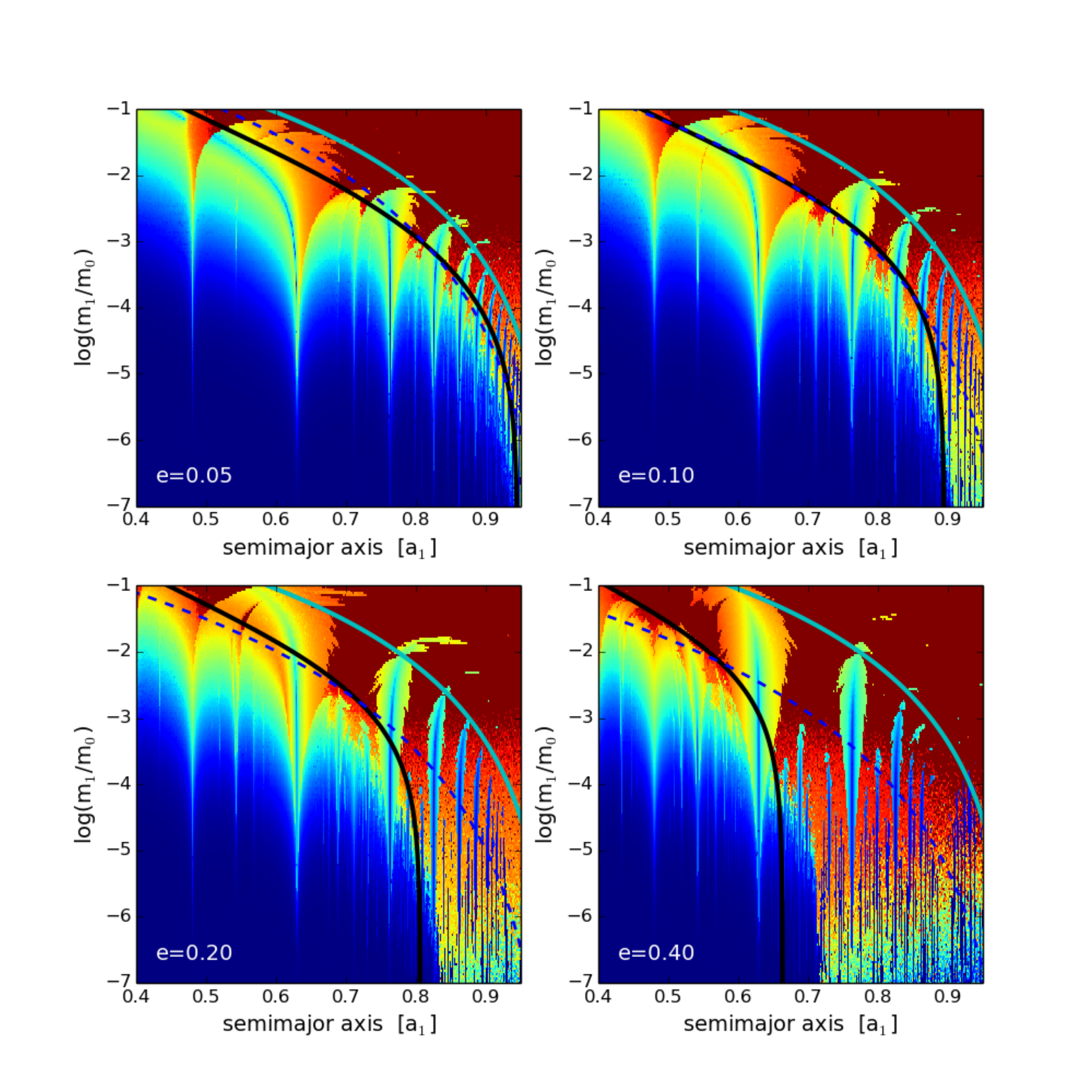}
\vspace{-3mm}
\caption{Dynamical maps of $\Delta e$ in the $(a/a_1,m_1/m_0)$ plane of initial conditions for four different values of the particles' eccentricities (indicated within each frame). Small changes in the eccentricity after $10^5$ years integration time are shown in blue, while unstable orbits leading to ejection are shown in red. In each plot, $a_{\rm unstable}$ in shown with a broad Cyan curve, while the Hill Stability limit (calculated for the specific value of $e$) is shown in black. The dashed blue line shows the analytical derivation of the overlap criterion for eccentric orbits developed by Deck et al. (2013).}
\label{fig11}
\end{figure}

Figure \ref{fig11} now shows four dynamical maps of $\Delta e$ where, contrary to Figure \ref{fig8}, we have not highlighted orbits that are unstable while maintaining low eccentricities. Each frame corresponds to a different initial eccentricity of the particle, as indicated in white text. As the eccentricity increases, so does the instability region close to the planet, even for small perturbing masses. The libration width of each resonance also increases, generating islands of very regular motion immersed in regions of highly chaotic behavior. Many high-degree MMRs are noticeable for $a \rightarrow a_1$, as well as for semimajor axis below that corresponding to the 2/1 commensurability. The resonance structure thus becomes increasingly complex, much more so than observed for circular orbits. 

The broad black curve indicates the Hill Stability limit, as calculated for each value of the eccentricity. We can see that it follows very closely the inner edge of the chaotic domain, and it appears fairly clear that all initial conditions with $a < a_{\rm Hill}$ are stable, at least for the total integration time (again set at $T=10^5$ orbits of $m_1$). The broad Cyan curves correspond to the empirical value of $a_{\rm unstable}$ developed in this work, as calculated for circular orbits. Surprisingly, the same functional form shows a very good agreement with the beginning of the global chaotic region, independently of the eccentricity of the particle. Thus, it appears that this upper stability limit is an extremely weak function of $e$, even up to $e=0.4$. 

Finally, the dashed blue line shows the predictions of the overlap criterion developed for eccentric orbits by Mustill and Wyatt (2012) and later refined by Deck et al. (2013). Using a simple analytical model for the growth of the libration domain as function of the eccentricity, both papers deduce that the overlap limit has a functional form proportional to $(e (m_1/m_0))^{1/5}$, although they differ in the value of the numerical constant. Here we have adopted the version of Deck et al. (2013). 

\begin{figure}[t!]
\centering
\includegraphics[width=13cm]{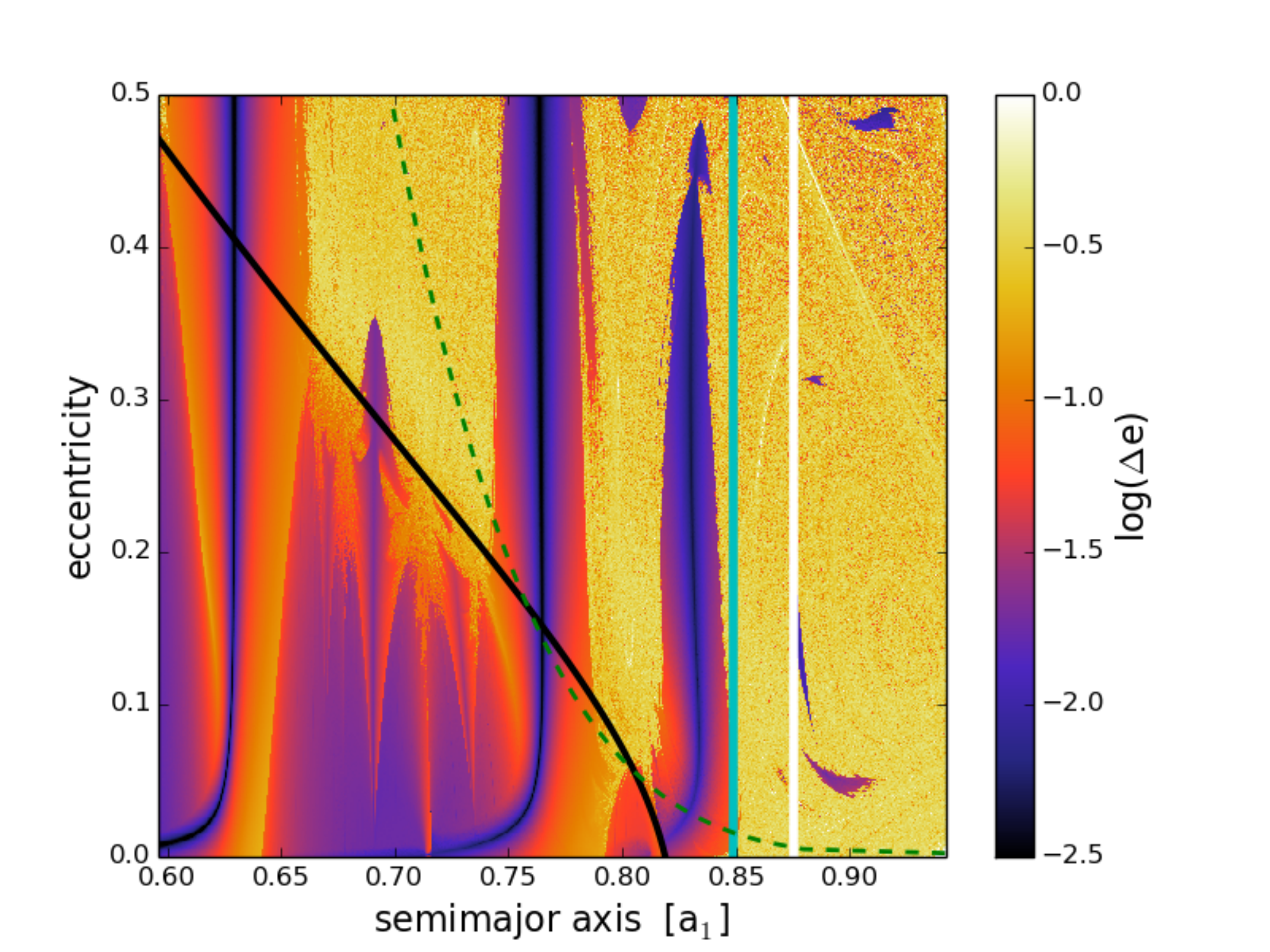}
\vspace{-3mm}
\caption{Dynamical map of $\Delta e$ for a grid of initial conditions in the $(a/a_1,e)$ plane of osculating elements (all angles equal to zero), with Jupiter as the perturber. Integration time was set to $T=10^3$ orbits of the perturber. The broad continuous curves show the Hill stability limit (black), the overlap limit deduced in this paper (cyan) and the value of $a_{\rm unstable}$ (white). The dashed green curve is the prediction of the eccentric overlap criterion by Deck et al. (2013).}
\label{fig12}
\end{figure}

Trying to extend the overlap criterion to non-circular orbits is complex. As we showed previously, the outer separatrix of first-order resonance does not exist for $e \sim 0$ but does appear for some minimum value, and thus should be taken into consideration at some point. Third-order resonances also become noticeable for eccentric orbits and begin to overlap with lower-order commensurabilities for sufficiently high eccentricities. 

Since it is difficult to take into consideration all these factors, it is not surprising that the analytical model does not show a good agreement with the dynamical maps. In fact, they do appear to be an ``average'' between both the circular resonance overlap and Hill stability limits. Curiously, however, for eccentricities $e \le 0.1$ there is a very good coincidence with the Hill Stability curve.

Figure \ref{fig12} shows a different example. This time we constructed a $\Delta e$ dynamical map in the $(a/a_1,e)$ representative plane, considering Jupiter (current mass and circular orbit) as the perturber. We defined a grid of initial conditions with $2048$ values of semimajor axis and $576$ values of the initial eccentricity. All angles were taken equal to zero. The total integration time was only $T=10^3$ orbits of the perturber, not sufficient to eject many unstable orbits, but sufficient to show large increases in the eccentricity. 

\begin{figure}[t!]
\centering
\includegraphics[width=12cm]{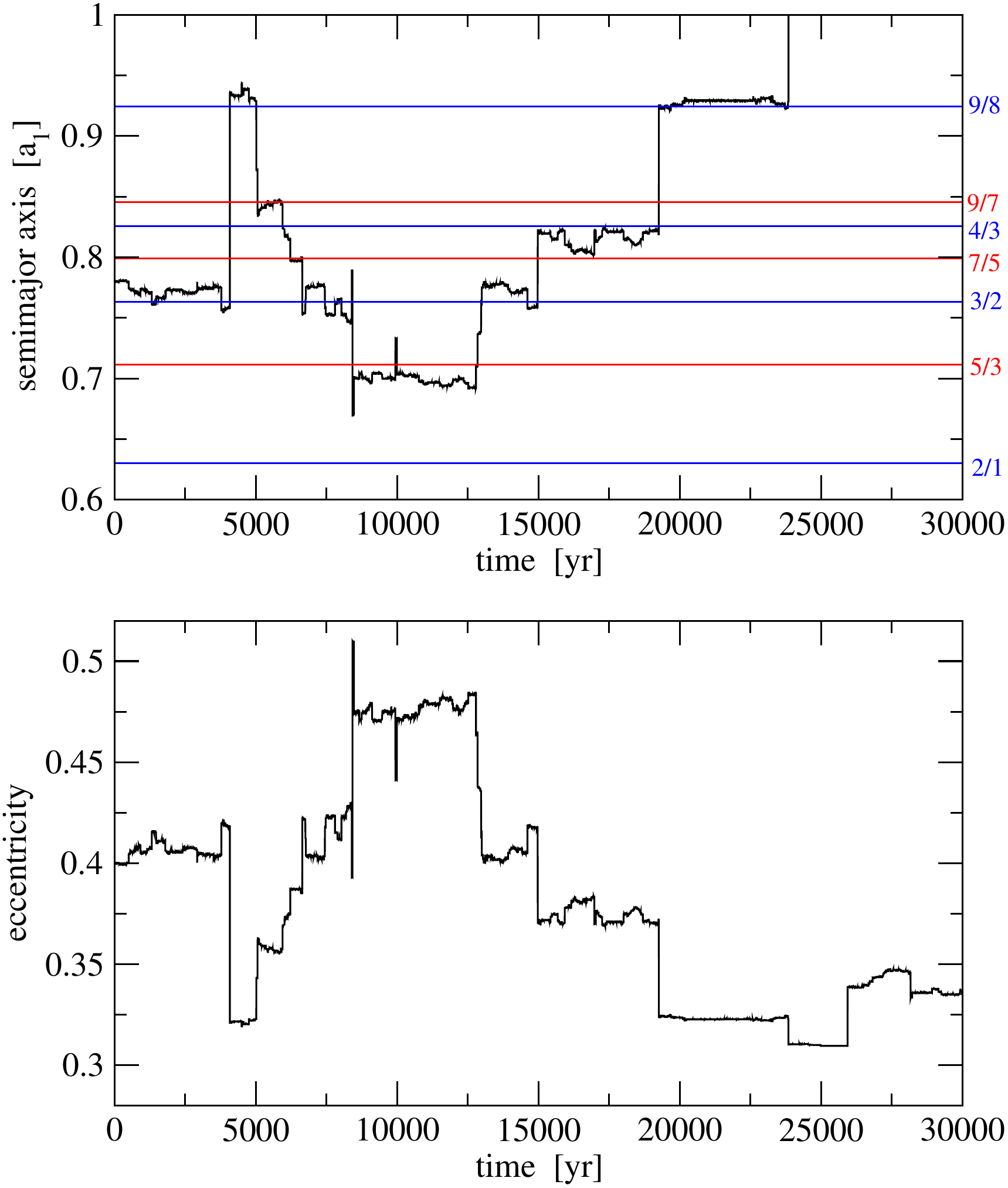}
\vspace{-3mm}
\caption{Numerical integration of a particle with $a = 0.8$, $e=0.4$ and all angles initially set to zero. The perturbing mass was chosen equal to $m_1/m_0 = 10^{-6}$. The particle exhibits a random walk between adjacent first and second-order MMRs before finally exceeding the perturber's orbit at $t \simeq 2.5 \times 10^4$ years. During the complete integration the eccentricity remains close to its initial value with $\Delta e$ of the order of $0.1$.}
\label{fig13}
\end{figure}

\begin{figure}[t!]
\centering
\includegraphics[width=13cm]{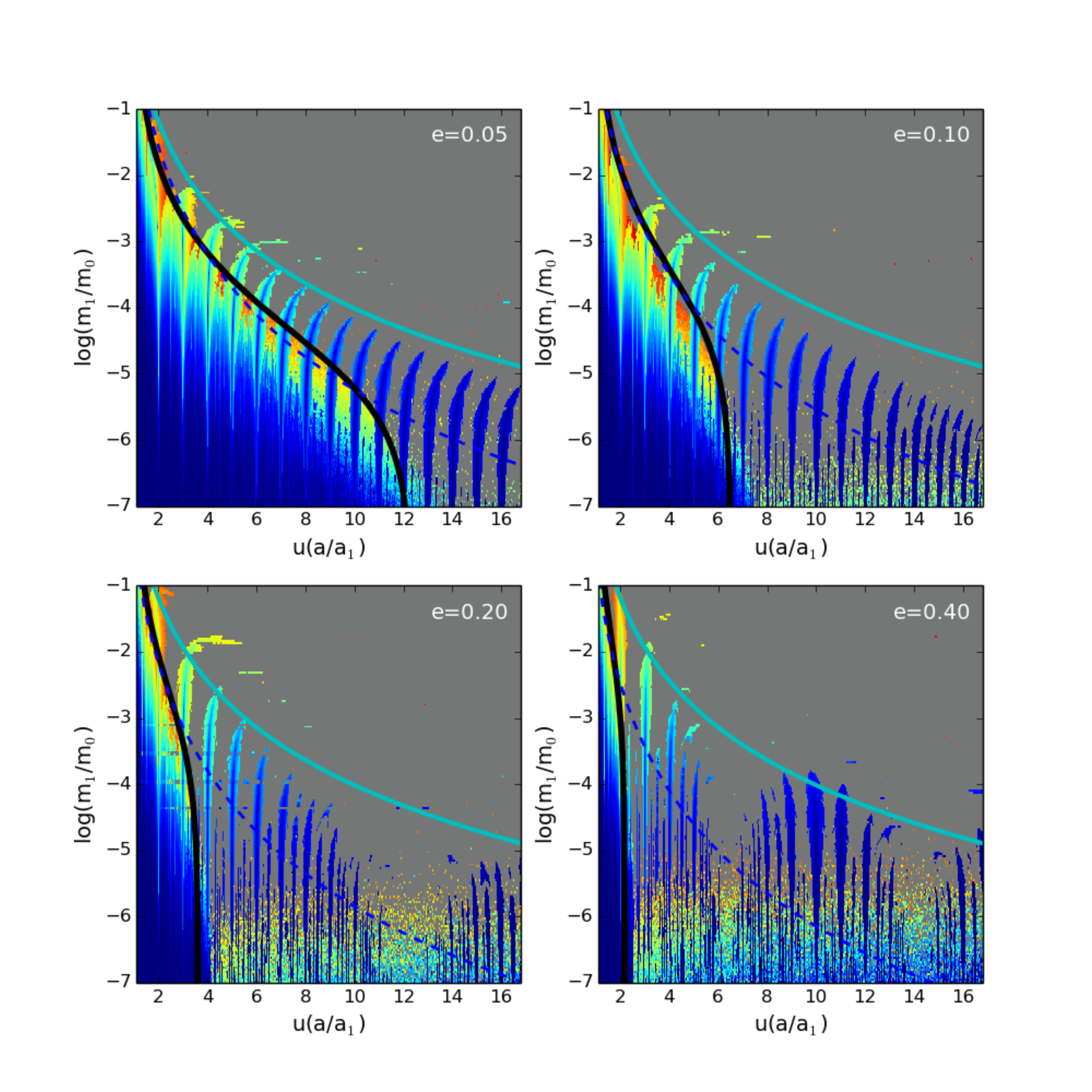}
\vspace{-3mm}
\caption{Similar to Figure \ref{fig11}, but now the abscissa is a function of the semimajor axis ratio defined by: $u(a/a_1) = -[(a/a_1)^{3/2}-1)]^{-1}$. In this new variable all the first-order resonances appear equidistant, thus allowing a better visualization of the resonant structure closer to the perturber. As before, the black curve shows the Hill stability limit, while cyan corresponds to $a_{\rm unstable}$. The eccentric overlap criterion of Deck et al. (2013) is shown as a dashed blue curve.}
\label{fig14}
\end{figure}

Apart from the main first-order commensurabilities (2/1, 3/2 and 4/3), there is distinct evidence of second and third-order resonances, particularly the 7/4 at $a/a_1 \simeq 0.69$ and the 8/5 at $a/a_1 \simeq 0.73$. These are negligible for near-circular orbits, but become increasingly important for higher eccentricities (e.g. Bodman and Quillen 2014). At $e \sim 0.2$ these begin to overlap and generate a large chaotic region close to the Hill Stability curve, shown here again as a broad black curve. 

Gladman (1993) also reported, from the result of a few numerical simulations, that some orbits above the Hill stability curve may survive for long time-spans as long as they satisfy the Hill condition for circular orbits. As we show here, the relationship between instability and the Hill limit is more complex.

The broad cyan curve shows our version of the resonance overlap criterion, as deduced for circular orbits, while the white vertical line shows the value of $a_{\rm unstable}$. In particular, the value of $a_{\rm overlap}$ shows a very good agreement with the beginning of the global chaotic sea, and appears fairly independent of the eccentricity. Finally, the dashed green curve corresponds to the eccentric overlap criterion as proposed by Deck et al. (2013). 

Figures \ref{fig11} and \ref{fig12} show that the region between $a_{\rm Hill}$ and $a_{\rm unstable}$ grows with increasing value of $e$ and lower values of $m_1$. This transition region is characterized by a complex resonant structure and shows the existence of both stable and unstable orbits. Stable motion is usually found deep inside the libration domain of mean-motion resonances, while unstable orbits abound elsewhere. 

For perturbing masses below $m_1/m_0 \sim 10^4-10^3$, the results of Figure \ref{fig11} seem to indicate that particles with $a > a_{\rm Hill}$ are not very excited and remain with low-to-moderate eccentricities. To investigate this in more detail, Figure \ref{fig13} shows the dynamical evolution of a single initial condition for a total integration time of $3 \times 10^4$ years. The perturbing mass was chosen equal to $m_1/m_0 = 10^{-6}$. As mentioned in the caption, the body suffers several jumps in semimajor axis, becoming temporarily trapped between adjacent first and second-order mean-motion resonances. After approximately 25000 years the orbit finally surpasses the semimajor axis of the perturber and eventually diffuses to the outer regions of the system. The eccentricity, however, remains bounded throughout the integration, with a maximum increase (with respect to its initial value) of the order of $\Delta e \simeq 0.1$. 

Since orbital instability is not necessarily associated to hyperbolic orbits, Figure \ref{fig14} repeats the results of Figure \ref{fig11}, where the gray region now shows all orbits that became unstable within $T=10^5$ orbits, either reaching $e >1$ or becoming planet crossers. We also plotted the data in terms of an auxiliary function $u(a/a_1)$ instead of the semimajor axis. This function is defined by:
\be
u(a/a_1) = - \biggl[ \biggl( \frac{a}{a_1} \biggr)^{3/2} - 1 \biggr]^{-1} .
\label{eq56}
\ee
In this new variable all first-order resonances appear equidistant, thus allowing us to have a better visualization of the structure of the representative plane closer to the perturbing mass. 

As before, the lower bound of the chaotic sea is very well represented by the circular estimate of $a_{\rm unstable}$, while the Hill Stability correctly delimits the appearance of the resonance forest. The eccentric overlap formula of Deck et al. (2013) is seen in dashed blue lines. As before, this later criterion shows a good agreement with the Hill limit for low eccentricities and planetary masses above $m_1/m_0 \sim 10^{-4}$, but rapidly diverges for both lower masses and higher eccentricities.

\section{Conclusions}

In this paper we presented the results of a series of high-resolution $\Delta e$ dynamical maps in a representative plane of the planar circular restricted three-body problem, whose aims where two-fold: (i) obtain a detailed visualization of the limit between stable and unstable orbits (in the Hill sense) and (ii) estimate the resonant structure of mean-motion commensurabilities and their relation with the stability limit. These results were used to test the predictions of the Hill stability limit and different versions of the resonance overlap criterion (circular and elliptic case).

In particular, using the Second Fundamental Model for Resonance (Henrard and Lema\^{\i}tre 1983), we obtained an alternative derivation of the overlap criterion based on the interaction between first and second-order commensurabilities which appears to be responsible for the transition between local and global chaotic motion. The resulting expression is reminiscent of the one presented by Wisdom (1980), although with a smaller numerical coefficient and slightly different functional dependence on the perturbing mass.

Chirikov's postulation of the overlap criterion was constructed for systems in which the interacting resonances occurred in the same set of canonical variables and, implicitly, with the same expressions for the integrals of motion. These conditions do not apply for the problem at hand where each mean-motion resonance is characterized by distinct expressions of the integral $N^*$ and, consequently, different set of canonical variables. Classical studies have circumvented this problem by avoiding canonical variables and analyzing their interaction in the $(a,e)$ plane, variables which are common to all MMRs. However, as we was seen throughout this paper, these orbital elements are not adequate and the outer separatrix does not appear for quasi-circular motion, leading to a series of assumptions that have not always been justified. Thus, although Chirikov's overlap criterion has been widely used in the circular restricted three-body problem, its applicability has not always been adequately verified.

It is important to stress that the aim of this paper has not been to introduce a new rigorous theory for resonance overlap, but to present a qualitative criteria that preserves the principles of the classical formulation but (more importantly) reproduces the dynamical features observed in the numerical simulations.

Finally, we showed that it is not possible to characterize the stability limit using a single criterion. The Hill Stability criterion defines the maximum value of the semimajor axis (for a given eccentricity) for which all initial conditions are (Hill) stable. On the other hand, an empirical variation of our resonance overlap limit (\ref{eq49}) is a useful proxy to estimate the minimum semimajor axis for which all initial conditions are unstable. In between lies a transition domain characterized by many resonant islands and rich in both stable and unstable motion. The size of this region grows with increasing eccentricity or lower values of the perturbing mass, and may occupy a significant portion of the phase space.

\begin{acknowledgements}
This work has been supported by research grants from Secyt/UNC, CONICET and FONCyT. The authors wish to thank IATE and the UNC for extensive use of their computing facilities, and to Pablo Benitez-Llambay for his assistance in developing the graphical routines. Finally, we would also like to express our gratitude to an anonymous referee for stimulating discussions that helped improve this work.
\end{acknowledgements}

\end{document}